\documentclass[10pt]{iopart}
\usepackage{graphicx,pstricks}
\usepackage{tikz}
\usepackage{sistyle}
\usepackage{amsfonts,iopams}
\usepackage{color,colortbl}
\usepackage{hyperref}
\usepackage{caption}
\usepackage{subcaption}

\newcommand{\Z}{{\mathbb{Z}}}
\usepackage{epsfig,epsf,amsfonts,amscd}

\begin{document}
%\sloppy
%\draft
\title{Flat and almost flat bands in the quasi-one-dimensional
Josephson junction array
}
\author{Daryna Bukatova and Yaroslav Zolotaryuk}
%\address{$^1$ 
%KAU}
\address
{Bogolyubov Institute for Theoretical Physics, National Academy of
Sciences of Ukraine, Kyiv 03143, Ukraine}
%\author{Yaroslav Zolotaryuk}
\ead{daryna.bukatova@gmail.com and yzolo@bitp.kiev.ua (corresponding author)}

%\affiliation{Bogolyubov Institute for Theoretical Physics, National Academy of
%Sciences of Ukraine, Kyiv 03680, Ukraine}
\date{\today}

\begin{abstract}
The dispersion law for the linear waves in the quasi-one-dimensional
array of inductively coupled Josephson junctions (JJ) is derived. The array 
has a multiladder structure that consists of the finite
number of rows ($N\ge 2$) in $Y$ direction  
and is infinite in $X$ direction. The spectrum of the linear
waves (Josephson plasmons) consists of $2N-1$ branches. Among these branches
there is a $N$-fold completely flat degenerate one that coincides with the 
Josephson plasma frequency. The remaining $N-1$ branches have a standard
Josephson plasmon dispersion law typical for 1D JJ arrays. 
Application of the uniform dc bias on the top of each 
vertical column of junctions
lifts the degeneracy and only one flat branch remains unchanged. The
rest of the previously flat branches become weakly dispersive. 
The parameter range where the flatness of these branches is maximal
has been discussed.
\end{abstract}

%\maketitle
% Uncomment for PACS numbers
%\pacs{\textcolor{red}{CHECk! 00.00, xx.xx, yy.yy}}
%
% Uncomment for keywords
\vspace{2pc}
\noindent{\it Keywords}: Weak superconductivity, Josephson junctions, Josephson junction arrays, dispersion, plasmons, flat bands.
%
% Uncomment for Submitted to journal title message

\submitto{\JPC}
%####################################################################
\section{Introduction}
\label{sec1}
%####################################################################

The concept of flat bands (FB) appears in  different fields of modern 
physics \cite{laf18aip}. A
FB in a spectrum (either energy or phonon) which 
is a completely dispersionless band 
$\mathbf{\nabla}_\mathbf{q}\,\omega(\mathbf{q})=0$ where
$\omega(\mathbf{q})$ is the dispersion law of the respective
system. 
The concept of FBs was introduced at first theoretically
for electron states in the two-dimensional dice lattice model \cite{s86prb}
and, some years later, for the itinerant electrons in the Hubbard model 
\cite{l89prl}. 
It was later extended to magnetic systems
\cite{dhr07prb}, Josephson junctions (JJs) 
\cite{phbgpp08prb,af19jpa} and Dirac materials \cite{hv11jetp,ggo21prb}. 
A remarkable consequence of the absence of dispersion is the existence
of localized states without breaking the translational invariance of the lattice. 
Although the first FB models appeared in the late 1980s, 
only about 10-15 years 
ago it became possible to observe FBs in different systems and to construct artificial FB systems. For example, several lattice 
structures with a FB have been realized in photonic lattices (such as waveguide arrays) \cite{vcm-irm-cwsm15prl} and cold atomic gases in optical lattices \cite{toinnt15sci}. 

The Josephson transmission lines (JTL) or Josephson junction arrays (JJAs) 
are examples of artificial superconducting systems that can support
FBs. For example, the frustrated JJA with FBs has been studied in
\cite{af19jpa}. It should be mentioned that JJAs are widely
studied for applications in quantum computing, for example, possibility
 of application of JTLs for quantum states readout was discussed 
 theoretically in \cite{ars06prb} and studied experimentally later 
 in \cite{fswbu14prl}. Since there has 
been active research conducted on propagation of current pulses through JTLs
\cite{bkrs19jpcs}, it is important to study their dispersion relations.

Ladders of JJs have been actively investigated due to observation of
various nonlinear phenomena such as vortex propagation \cite{yls93prb,ldhhbb94prb},
meandering \cite{acfflu99prl} and, finally, prediction \cite{fmmfa96epl} and
experimental discovery \cite{tmo00prl,baufz00prl,bau00pre} of discrete
breathers \cite{fw98pr}. 
It should be noted that while Josephson vortices can exist in the standard
1D JJA, discrete breathers require at least a simple JJ ladder. 
In \cite{baufz00prl,mffzp01pre} it was shown that the Josephson plasmon 
 spectrum for the
anisotropic Josephson junction ladder (JJL) with two horizontal rows 
has a flat band. It is natural to generalize this ladder into a quasi-one-dimensional 
ladder-like structure with an arbitrary number of horizontal rows $N$. 
We will call it an $N$-row JJA or a 
quasi-one-dimensional Josephson junction array (Q1D JJA). 

The aim of this work is to find and analyse the plasmon spectrum 
and the the corresponding wave amplitudes 
of the Q1D JJA with the arbitrary number of rows, 
and, in particular, to establish whether this spectrum still has a 
FB for any number of rows as it does for the 2-row case.

This paper is organized as follows. In the next section we present the equations
of motion for the Q1D JJA with $N$ rows. In the third section the dispersion law for
an arbitrary number of rows is obtained and its main properties are 
discussed. Outlines and conclusions are given in the last section.

%####################################################################
\section{Equations of motion}
\label{sec2}
%####################################################################

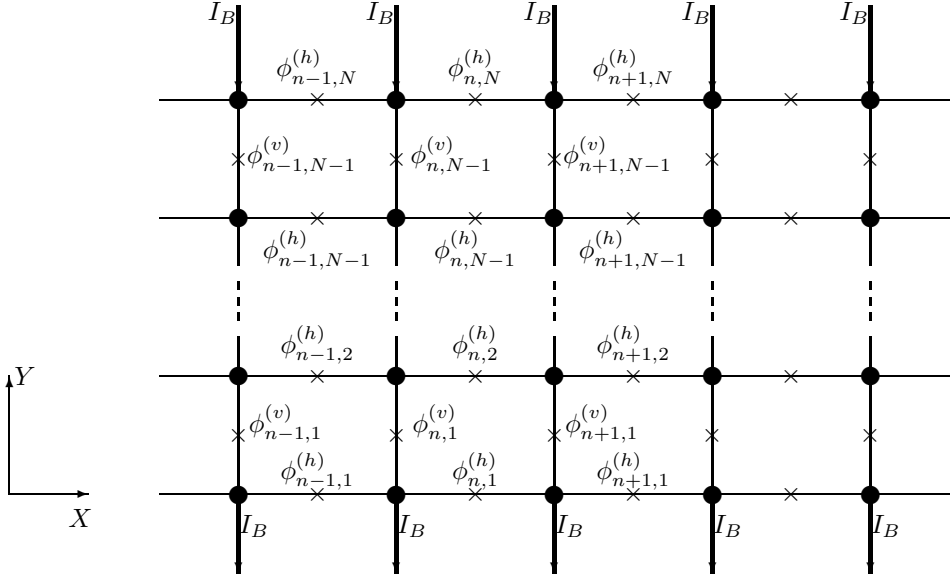
\begin{figure}[htb]
%	\centering{\includegraphics[width=0.48\linewidth]{fig1.eps}}
\setlength{\unitlength}{10.5cm}
\begin{picture}(1.,0.73)
%\thicklines
%\matrixput(0.2, 0.0){4}(0.2, 0.15){2}{\circle*{0.025}}
\multiput(0.3, 0.1)(0.2, 0.0){5}{\circle*{0.025}}
\multiput(0.3, 0.25)(0.2, 0.0){5}{\circle*{0.025}}
\multiput(0.3, 0.45)(0.2, 0.0){5}{\circle*{0.025}}
\multiput(0.3, 0.6)(0.2, 0.0){5}{\circle*{0.025}}

\multiput(0.4, 0.1)(0.2,0){4}{\makebox(0,0){$\times$}}
\multiput(0.4, 0.25)(0.2,0){4}{\makebox(0,0){$\times$}}
\multiput(0.4, 0.45)(0.2,0){4}{\makebox(0,0){$\times$}}
\multiput(0.4, 0.6)(0.2,0){4}{\makebox(0,0){$\times$}}
\multiput(0.3, 0.175)(0.2,0.){5}{\makebox(0,0){$\times$}}
\multiput(0.3, 0.525)(0.2,0.){5}{\makebox(0,0){$\times$}}

\multiput(0.3, 0.32)(0,0.02){3}{\line(0,1){0.010}}
\multiput(0.5, 0.32)(0,0.02){3}{\line(0,1){0.010}}
\multiput(0.7, 0.32)(0,0.02){3}{\line(0,1){0.010}}
\multiput(0.9, 0.32)(0,0.02){3}{\line(0,1){0.010}}
\multiput(1.1, 0.32)(0,0.02){3}{\line(0,1){0.010}}
%\draw[dashed] (10,1) -- (10,11);
%\tikzstyle{dotted}=[dash pattern=on \pgflinewidth off 2pt]

%\protect\tikz[baseline]{\protect\draw[line width=0.2mm,densely dashed] (0.3,.1)--++(0.3,0.3) ;}
%\put(0, 0){\vector(1, 0){0.5}}
\multiput(0.3,0)(0.2,0){5}{\line(0,1){0.3}}
\multiput(0.3,0.39)(0.2,0){5}{\line(0,1){0.3}}

\put(0.2, 0.1){\line(1, 0){1.0}}
\put(0.2, 0.25){\line(1, 0){1.0}}
%\put(0.2, 0.2){\line(1, 0){1.0}}
%\put(0.2, 0.35){\line(1, 0){1.0}}
\put(0.2, 0.45){\line(1, 0){1.0}}
\put(0.2, 0.6){\line(1, 0){1.0}}

%\linethickness{0.5pt}

\put(0.38, 0.53){\makebox(0,0){$ \phi^{(v)}_{n-1,N-1}$}}
\put(0.57, 0.53){\makebox(0,0){$\phi^{(v)}_{n,N-1}$}}
\put(0.78, 0.53){\makebox(0,0){$\phi^{(v)}_{n+1,N-1}$}}
\put(0.4, 0.64){\makebox(0,0){$\phi^{(h)}_{n-1,N}$}}
\put(0.6, 0.64){\makebox(0,0){$\phi^{(h)}_{n,N}$}}
\put(0.8, 0.64){\makebox(0,0){$\phi^{(h)}_{n+1,N}$}}

\put(0.4, 0.41){\makebox(0,0){$\phi^{(h)}_{n-1,N-1}$}}
\put(0.6, 0.41){\makebox(0,0){$\phi^{(h)}_{n,N-1}$}}
\put(0.8, 0.41){\makebox(0,0){$\phi^{(h)}_{n+1,N-1}$}}

\put(0.4, 0.29){\makebox(0,0){$ \phi^{(h)}_{n-1,2}$}}
\put(0.6, 0.29){\makebox(0,0){$\phi^{(h)}_{n,2}$}}
\put(0.8, 0.29){\makebox(0,0){$\phi^{(h)}_{n+1,2}$}}

\put(0.36, 0.19){\makebox(0,0){$\phi^{(v)}_{n-1,1}$}}
\put(0.55, 0.19){\makebox(0,0){$\phi^{(v)}_{n,1}$}}
\put(0.76, 0.19){\makebox(0,0){$ \phi^{(v)}_{n+1,1}$}}

\put(0.4, 0.13){\makebox(0,0){$ \phi^{(h)}_{n-1,1}$}}
\put(0.6, 0.13){\makebox(0,0){$ \phi^{(h)}_{n,1}$}}
\put(0.8, 0.13){\makebox(0,0){$\phi^{(h)}_{n+1,1}$}}

\put(0.01,0.1){\vector(1,0){0.1}}
\put(0.01,0.1){\vector(0,1){0.15}}
\put(0.1,0.07){\makebox(0,0){$ X$}}
\put(0.03,0.25){\makebox(0,0){$ Y$}}
\linethickness{1.5pt}
\multiput(0.3, 0.1)(0.2,0){5}{\vector(0,-1){0.10}}
\multiput(0.3, 0.72)(0.2,0){5}{\vector(0,-1){0.11}}
\multiput(0.32,0.06)(0.2,0){5}{\makebox(0,0){$ I_B$}}
\multiput(0.28,0.71)(0.2,0){5}{\makebox(0,0){$ I_B$}}
%\put(0.1,0.1){\vector(1,0){0.1}}
%\put(0.3,0.1){\vector(1,0){0.1}}
\end{picture}
	\caption{Schematic view of the Q1D JJA with $N$ rows. 
	Only two top and two bottom rows are
	shown. The crosses $\times$ denote
	locations of the respective junctions. Thick arrows show the
	direction and location of the incoming and outcoming dc bias $I_B$.}
\label{fig1}
\end{figure}

We consider a ladder-like array of JJs that consists of
a finite number ($N$) of rows in Y direction as shown in figure \ref{fig1}.
The length of each of the rows in the X direction is supposed to be 
much greater than $N$. Since we are interested in the plane waves of the
array we may assume it to be infinite. 
%@@@@@@@@@@@@@@@@@@@@@@@@@@@@@@@@@@@@@@@@@@@@@@@@@@@@@@@@@@@@@@@@@@@@
%@@@@@@@@@@@@@@@@@@@@@@@@@@@@@@@@@@@@@@@@@@@@@@@@@@@@@@@@@@@@@@@@@@@@
The respective phase of the junction will be denoted as $\phi^{(v,h)}_{n,k}$
where the superscript $v$ or $h$ will denote whether the junction 
belongs to the row (a horizontal one) or to the bridge (a vertical one). The
set of subscripts $(n,k)$ enumerates the junction along the 
$X$ and $Y$ axis, respectively. 
The array is uniformly biased by the
 dc current $I_B$ at each %$\phi^{(v,h)}_{n,N}$
$(n,N)$th junction and
 the same current is extracted from the each% $\phi^{(v,h)}_{n,1}$
$(n,N)$th junction.
The equations for the time evolution of the Josephson phases for each
of the junctions within the resistively and capacitatively
shunted (RCSJ) model \cite{kkl86} are given 
by the following set of equations
%--------------------------------1------------------------------------
\begin{eqnarray}\nonumber
&&\frac{C_{v,h}\hbar}{2e}\frac{d^2}{dt^2}\phi^{(v,h)}_{n,k} +
\frac{\hbar}{2eR_{v,h}}\frac{d}{dt} {\phi}^{(v,h)}_{n,k}+
I_c^{(v,h)}\sin \phi^{(v,h)}_{n,k}= I^{(v,h)}_{n,k},\;\;\\
&&n \in \Z,\; v: k=\overline{1,N-1}, \; h: k=\overline{1, N}.
\label{1}
\end{eqnarray}
%---------------------------------------------------------------------
Here $C_{v,h}$ is the capacitance of the vertical or horizontal 
junction, $R_{v,h}$ is its resistance and $I^{(v,h)}_{c}$ is
its critical current, respectively. Finally, $I^{(v,h)}_{n,k}$ is the current
that flows through the $(n,k)$th junction. 

For the derivation of the evolution equations one can consult 
paper \cite{bw98prb}. Here only the main points of the derivation 
will be repeated.
For each of these currents we need to write the Kirchhoff's equations
and the flux quantization law. The latter connects the mesh currents
$I^{(m)}_{n,k}$ in the cell formed by the vertical junctions $(n,k)$
and $(n+1,k)$ and horizontal junctions $(n,k)$ and $(n,k+1)$ 
and the magnetic flux through this cell: 
%--------------------------------------------------------------------
\begin{equation}\label{JJ:5}
I^{(m)}_{n,k}=-\frac{\Phi_{n,k}}{L}=-{\Phi_0\over {2\pi L}}\left 
(\phi^{(v)}_{n+1,k}-\phi^{(v)}_{n,k}+\phi^{(h)}_{n,k+1}-\phi^{(h)}_{n,k}
\right).
\end{equation}
%--------------------------------------------------------------------
Here $L$ is the self-inductance of the cell and $\Phi_0=\pi \hbar/e$
is the magnetic flux quantum.
It is convenient to introduce the dimensionless variables in the
following way:
%--------------------------------------------------------------------
\begin{equation}
\tau=\omega_p t,\, \gamma=\frac{I_B}{I_c^{(v)}}, \,
\beta_L=\frac{2\pi I_c^{(v)} L}{\Phi_0},\, 
\eta=\frac{I_C^{(h)}}{I_C^{(v)}}=\frac{C_h}{C_v}=\frac{R_v}{R_h}.
\end{equation}
%--------------------------------------------------------------------
The coupling constant
$\beta_L$ measures the discreteness of the array. The dimensionless 
dissipation parameter is then
%$\alpha=\Phi_0/(2\pi I_c^{(v)} R_v)$
$\alpha=\Phi_0 \omega_p/(2\pi I_c^{(v)} R_v)$, and the time is normalized to the inverse
Josephson 
plasma frequency $\omega_p^{-1}=\sqrt{C_v\Phi_0/(2\pi I_c^{(v)})}$. 
Finally, the parameter $\eta$ measures the anisotropy between the
junctions placed in the rows and columns of the array.
For the sake of convenience we also introduce the nonlinear operator
%--------------------------------------------------------------------
\begin{equation}
 {\cal N} (x) \equiv \ddot{x} + \alpha \dot{x} + \sin x .
\end{equation}
%--------------------------------------------------------------------
With the help of this operator the equations of motion are written 
inside the Q1D JJA
%--------------------------------------------------------------------
\begin{eqnarray}\nonumber
&{\cal N} (\phi^{(v)}_{n,k})=\gamma +  
\frac{1}{\beta_L} \left ( \hat{\Delta}_x \phi^{(v)}_{n,k} 
+\hat{\nabla}_x \phi^{(h)}_{n-1,k+1}-\hat{\nabla}_x \phi^{(h)}_{n-1,k}
\right ),\\
&\;\; n \in \Z~,\;k=\overline{1,N-1},\;  \label{8}\\ 
&{\cal N} (\phi^{(h)}_{n,k})= \frac{1}{\eta \beta_L} \left(\hat{\Delta}_y \phi^{(h)}_{n,k} +
\hat{\nabla}_x\phi^{(v)}_{n,k}-\hat{\nabla}_x\phi^{(v)}_{n,k-1} \right),
\nonumber \\
&  n \in \Z,~ \,k=\overline{2,N-1},  \label{9}
\end{eqnarray}
%--------------------------------------------------------------------
and on the border rows ($k=1,N$)
%--------------------------------------------------------------------
\begin{eqnarray}
&&{\cal N} (\phi^{(h)}_{n,1})= \frac{1}{\eta \beta_L}
\left ( \hat{\nabla}_x \phi^{(v)}_{n,1} + \hat{\nabla}_y 
\phi^{(h)}_{n,1} \right), \; n \in \Z~\label{10}\\
&&{\cal N} (\phi^{(h)}_{n,N})= -\frac{1}{\eta \beta_L}\left (
\hat{\nabla}_x \phi^{(v)}_{n,N-1}+\hat{\nabla}_y 
\phi^{(h)}_{n,N-1}\right ),\; n \in \Z~ \label{11}.
\end{eqnarray}
%--------------------------------------------------------------------
These equations represent a system of $2N-1$ coupled discrete sine-Gordon
equations. 
Here the difference operators introduced for the sake of simplicity
are given by the following expressions:
%--------------------------------------------------------------------
\begin{eqnarray}\label{11a}
&&{\hat \triangle}_x \phi_{n,k} \equiv \phi_{n+1,k} - 2 \phi_{n,k} + 
\phi_{n-1,k},\;\\
&&{\hat \triangle}_y \phi_{n,k} \equiv \phi_{n,k+1} - 2 \phi_{n,k} + 
\phi_{n,k-1},\; \nonumber \\
&&{\hat\nabla}_x\, \phi_{n} \equiv \phi_{n+1,k} - \phi_{n,k},\;
{\hat \nabla}_y\, \phi_{n,k} \equiv \phi_{n,k+1} - \phi_{n,k}.
\nonumber
\end{eqnarray}
%--------------------------------------------------------------------
In real JJAs the dimensionless dissipation parameter $\alpha$ is
rather small, $\alpha \lesssim 0.1$ \cite{u98pd}. Moreover, this
parameter provides us with information on how fast the Josephson 
plasmons decay. Our aim is to compute the plasmon spectrum, and, 
therefore, $\alpha$ will be neglected throughout the next sections.

%####################################################################
%####################################################################
\section{Plasmon bands and their properties}
\label{sec3}
%####################################################################
%####################################################################
%
%At this point we start deriving the dispersion law for the 
%Josephson plasmons and to study their properties as functions
% of the model parameters

%####################################################################
\subsection{Dispersion law derivation}
%####################################################################
In order to calculate the spectrum of the Josephson plasma waves 
one has to expand the equations of motion (\ref{8})-(\ref{11}) around 
the steady state
%------------------------------12------------------------------------
\begin{eqnarray} \label{12}
\phi^{(v)}_{n,k}=\arcsin \gamma,\;k=\overline{1,N-1};\; 
\phi^{(h)}_{n,k}=0, \;k=\overline{1,N},\; n\in \Z \, ,
\end{eqnarray}
%--------------------------------------------------------------------
that corresponds to the spatially uniform superconducting state of the
whole array. We will study linear waves that propagate in the $X$ direction.
The plane wave ansatz for the small deviations from
the steady state
%--------------------------------------------------------------------
\begin{eqnarray}\label{13}
\nonumber
&&\left (\delta \phi^{(v)}_{n,1},\cdots, \delta\phi^{(v)}_{n,N-1},
\delta \phi^{(h)}_{n,1},\cdots, \delta\phi^{(h)}_{n,N} \right )^T 
=\\ 
&&=\left (A^{(v)}_1, \cdots,  A^{(v)}_{N-1}, 
A^{(h)}_1, \cdots, A^{(h)}_{N}
\right)^T e^{\imath (q n + \omega \tau)} + \mbox{c.c}, 
\end{eqnarray}
%--------------------------------------------------------------------
is substituted into the equations of motion (\ref{8}-\ref{11}).
The resulting characteristic polynomial is given as 
a determinant of the respective $2N-1 \times 2N-1$ matrix. After
some calculations (see \ref{app-a} for details) the characteristic 
polynomial can be written explicitly and, after some manipulations, 
factorized as
%--------------------------------------------------------------------
\begin{equation}\label{12a}
\chi(\omega^2)=\frac{1}{\eta^{N-1}}(\omega^2-1)\prod_{n=2}^{N} 
[d_0-\alpha_n d_1],~~ \alpha_n=1-2\cos \frac{\pi (n-1)}{N},
\end{equation}  
%--------------------------------------------------------------------
 where 
%--------------------------------------------------------------------
\begin{eqnarray}
&&d_0(\omega^2)=\eta [\omega^2_i(q)-\omega^2] \left 
(1+\frac{1}{\eta\beta_L}-\omega^2\right )-\frac{2(1-\cos{q})}{\beta_L^2},  \\
&&d_1(\omega^2)=- \frac{1}{\beta_L} [\omega^2_i(q)-\omega^2]+
\frac{2(1-\cos{q})}{\beta_L^2},\\
&&\omega^2_i(q) \equiv \sqrt{1-\gamma ^2}+\frac{2}{\beta_L } (1-\cos q).
\end{eqnarray}
%--------------------------------------------------------------------
From here one concludes that there is one flat band with $\omega^2=1$.
The expression $\omega_i(q)$ is the dispersion law of the biased 
standard one-dimensional JJA where only the
 vertical Josephson junctions (see \cite{ucm93prb,wszo95prl}) are present.
The rest of the dispersion curves is derived from the equality
$d_0=\alpha_n d_1$. Thus, the whole set of the $2N-1$ dispersion branches
can be written as
%----------------------------16-17-----------------------------------
\begin{eqnarray}\label{16}
&\omega_0^2=1,\\
\label{17}
&\omega_{\pm n}^2(q)=\frac{1}{2}\left[1+\omega_i^2(q)+\frac{1+
\alpha_{n+1}}{\eta\beta_L} \right]
\pm \\
\nonumber
&\pm \sqrt{\frac{1}{4}\left[\omega_i^2-\left(1+\frac{1+\alpha_{n+1}}{\eta\beta_L} \right)\right ]^2+2(1+\alpha_{n+1})\frac{1-\cos q}{\eta\beta_L^2}},\\
%\pm \sqrt{\frac{\omega_i^4}{4}+\frac{1}{4}\left(1+\frac{1+\alpha_{n+1}}{\eta\beta_L} %\right)^2+2(1+\alpha_{n+1})\frac{1-\cos q}{\eta\beta_L^2}},\\
\nonumber
&n=\overline{1,N-1}\,. 
%&&\alpha_n=1-2\cos \frac{\pi (n-1)}{N}\;,n=1,\ldots,N-1,
\end{eqnarray}
%--------------------------------------------------------------------
The constant $\alpha_n$ is given in equation (\ref{12a}) and
of importance are the values for $n=\overline{2,N}$. 
For $N=2$ rows there is just one
value $\alpha_2=1$. For $N=3$ rows there are two values, $\alpha_2=0$,
$\alpha_3=1$. For larger $N$ the values of $\alpha_n$ will pack
the interval %$]-3,1[$ 
$]-1,3[$ more and more densely.

\begin{figure}[htb]
\centering
\begin{subfigure}{.33\textwidth}
 \centering
 \includegraphics[width=.99\linewidth]{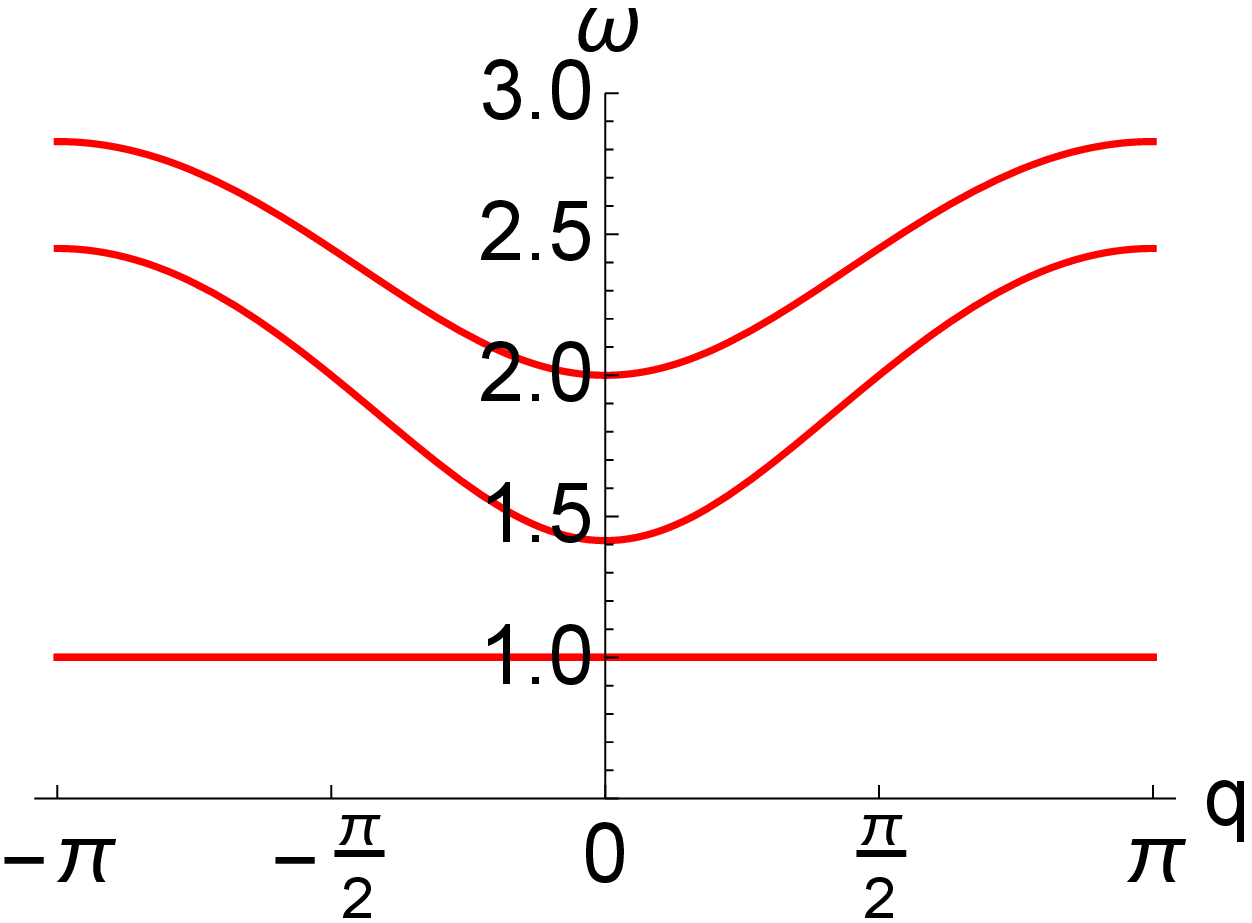}
 \caption{$\gamma = 0$}
 \label{fig2:sub1}
 \end{subfigure}
\begin{subfigure}{.33\textwidth}
 \centering
 \includegraphics[width=.99\linewidth]{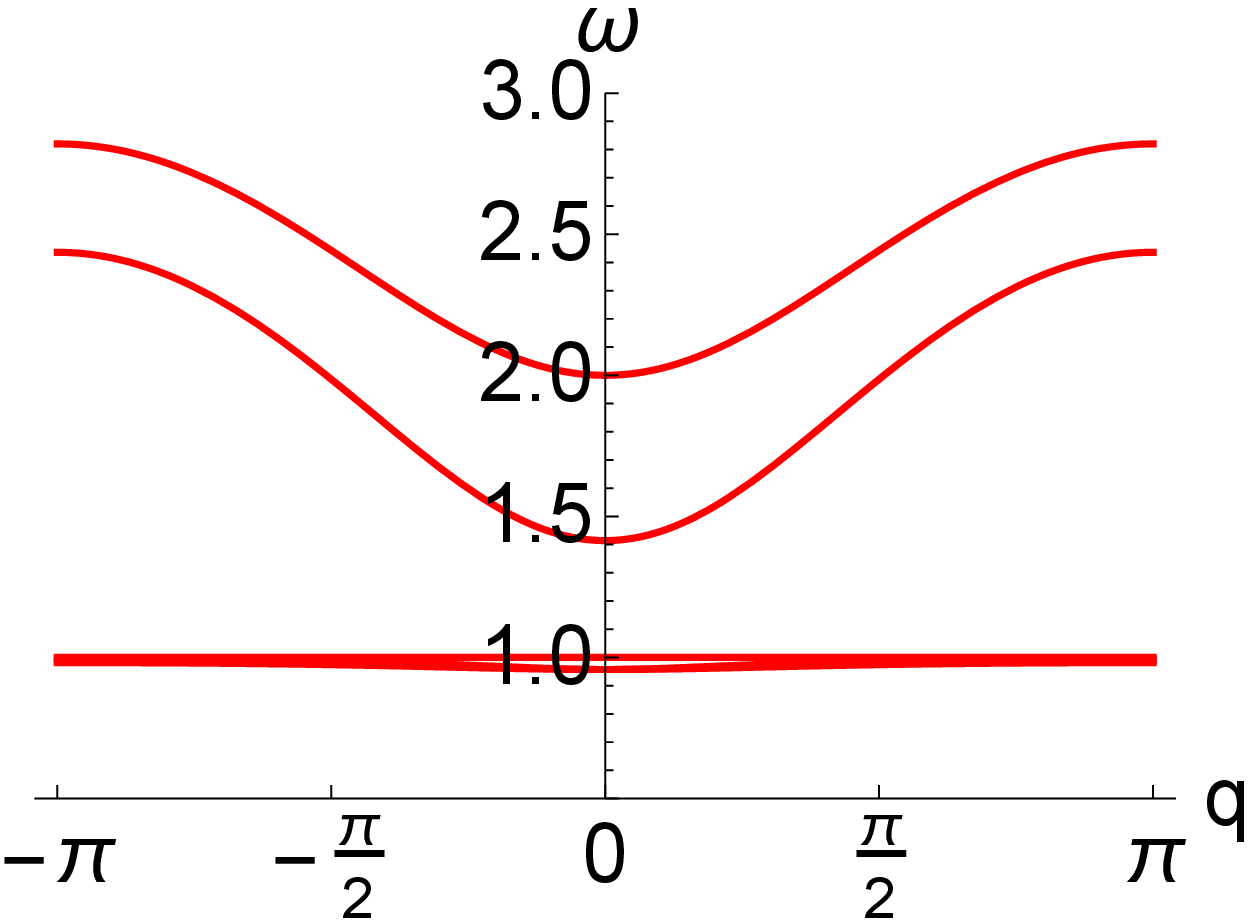}
 \caption{$\gamma = 0.4$}
 \label{fig2:sub2}
 \end{subfigure}%
 \begin{subfigure}{.33\textwidth}
 \centering
 \includegraphics[width=.99\linewidth]{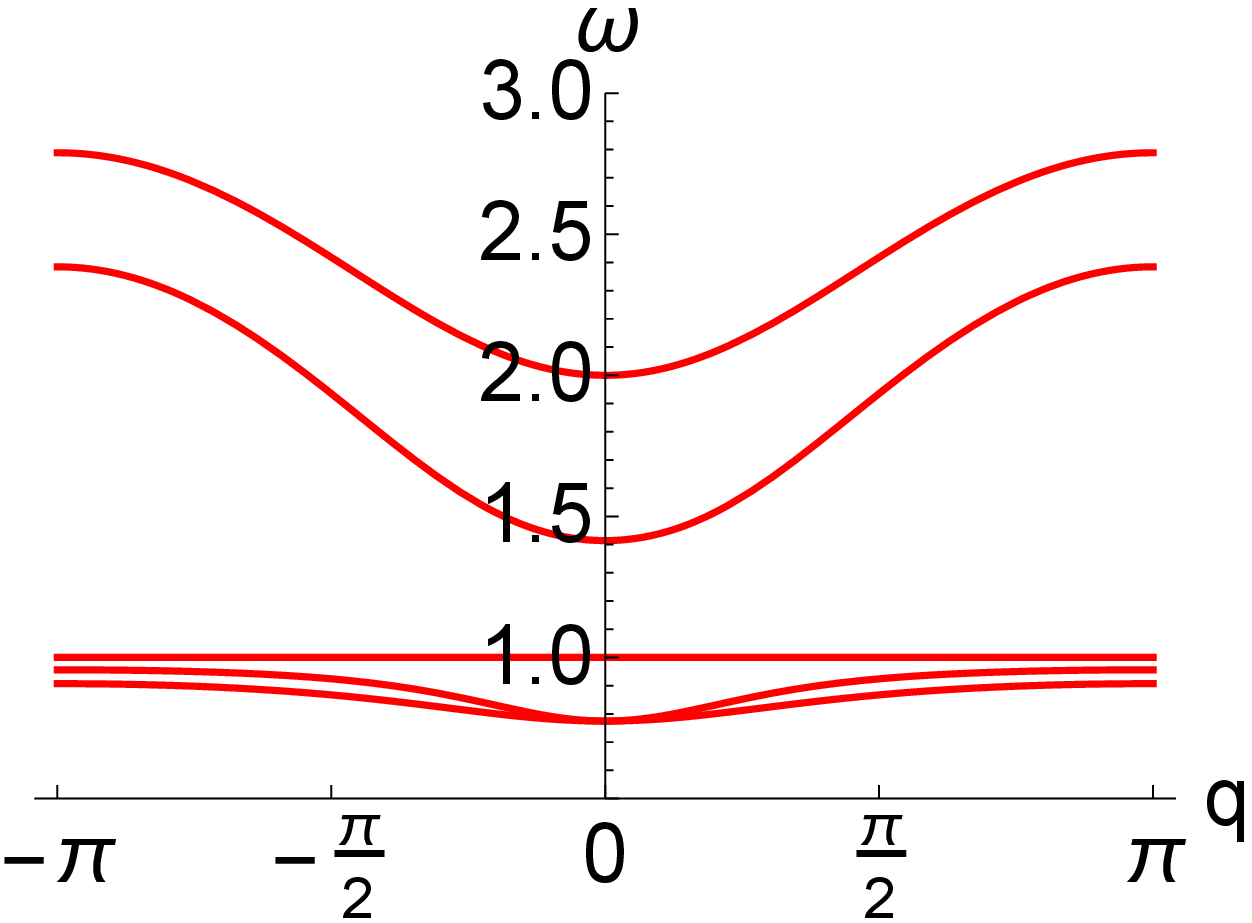}
 \caption{$\gamma = 0.8$}
 \label{fig2:sub3}
\end{subfigure}
\caption{Plasmon bands for $N=3$ row array, with 
$\eta=1,~\beta_L =1$ and different values of the dc bias $\gamma$.}
\label{fig2}
\end{figure}

For example, for the particular case of $N=3$ rows, the dispersion 
law consists of 5 branches.  
All of them are shown in figure \ref{fig2} for different values
of dc bias.
The branches are indexed in such a way that the dispersionless
branch $\omega_0$ is placed in the middle, the branches with 
positive subscript lie above the $\omega_0$ branch and the
branches with the negative subscript lie below. 
%@@@@@@@@@@@@@@@@@@@@@@@@@@@@@@@@@@@@@@@@@@@@@@@@@@@@@@@@@@@@@@@@@@@@
%@@@@@@@@@@@@@@@@@@@@@@@@@@@@@@@@@@@@@@@@@@@@@@@@@@@@@@@@@@@@@@@@@@@@
In the $N=2$ case the formulae (\ref{16}-\ref{17}) repeat the already known result \cite{mffzp01pre}.
Thus, generally speaking, the following inequality is valid for all
plasmon branches (\ref{16})-(\ref{17}): 
$\omega_{-N+1}<\cdots\omega_{-1}<\omega_0=1<\omega_1
<\cdots \omega_{N-1}$.
The set of branches with $\omega_{n>0}$ appear above the branch 
$\omega_0=1$. They are significantly dispersive and are well
separated from each other. The branches $\omega_{n<0}$ depend
crucially on the presence of the dc bias. When $\gamma=0$
they are all {\it degenerate} $\omega_{-N+1}=\cdots=\omega_{-1}=1$.
When the bias is applied the degeneracy is lifted. In that
case the branches with $n<0$ detach from the $\omega_0=1$ branch
and lie below it. If the bias is small these branches still
remain almost flat [see figure \ref{fig2}(b)].
Thus, in the general case of the dc biased Q1D array its plasmon
spectrum consists of
\begin{itemize}
\item one flat band with $\omega=1$;
\item $N-1$ strongly dispersive bands;
\item $N-1$ weakly dispersive bands that become completely
flat at $\gamma=0$.
\end{itemize}

%####################################################################
\subsection{Dispersion law properties for the different model parameters}
%####################################################################

In the long wave limit $q\to 0$ the frequencies of all modes
satisfy the following relations:
%--------------------------------------------------------------------
\begin{eqnarray}
\omega_n(0)=\left \{ \begin{array}{cc}
\sqrt{1+\frac{1+\alpha_{n+1}}{\eta \beta_L}},& n>0  \\
(1-\gamma^2)^{1/4},& n<0.
\end{array}
\right.
\end{eqnarray}
%--------------------------------------------------------------------
Thus, even when the degeneracy is lifted at $\gamma\neq 0$ the
lower branches remain degenerate at the point $q=0$ while
the upper branches are completely separated. In the limit of
strong discreteness ($\beta_L \gg 1$) the difference between
the neighboring branches reads $\Delta\omega_n(0)=\omega_{n+1}(0)-\omega_n(0)\approx 2\sin[\pi/(2N)]\sin[(2n+1)\pi/(2N)](\eta\beta_L)^{-1}
+{\cal O}[(\eta\beta_L)^{-2}]$.

\begin{figure}[htb]
	\centering
	\begin{subfigure}{.32\textwidth}
		\centering
		\includegraphics[width=.99\linewidth]{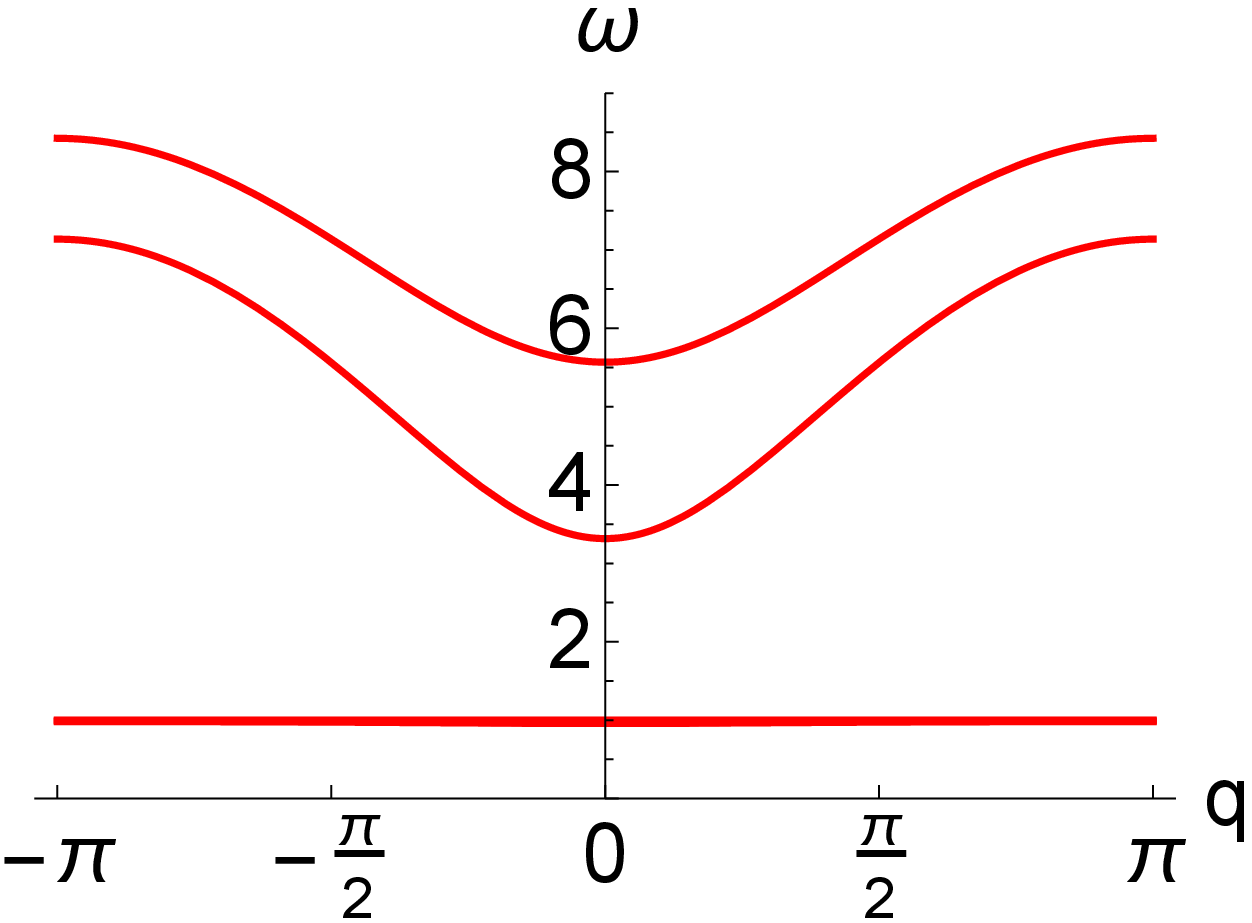}
		\caption{$ \beta_L = 0.1 $}
		\label{fig3a}
	\end{subfigure}
	\hskip1pt
	\begin{subfigure}{.32\textwidth}
		\centering
		\includegraphics[width=.99\linewidth]{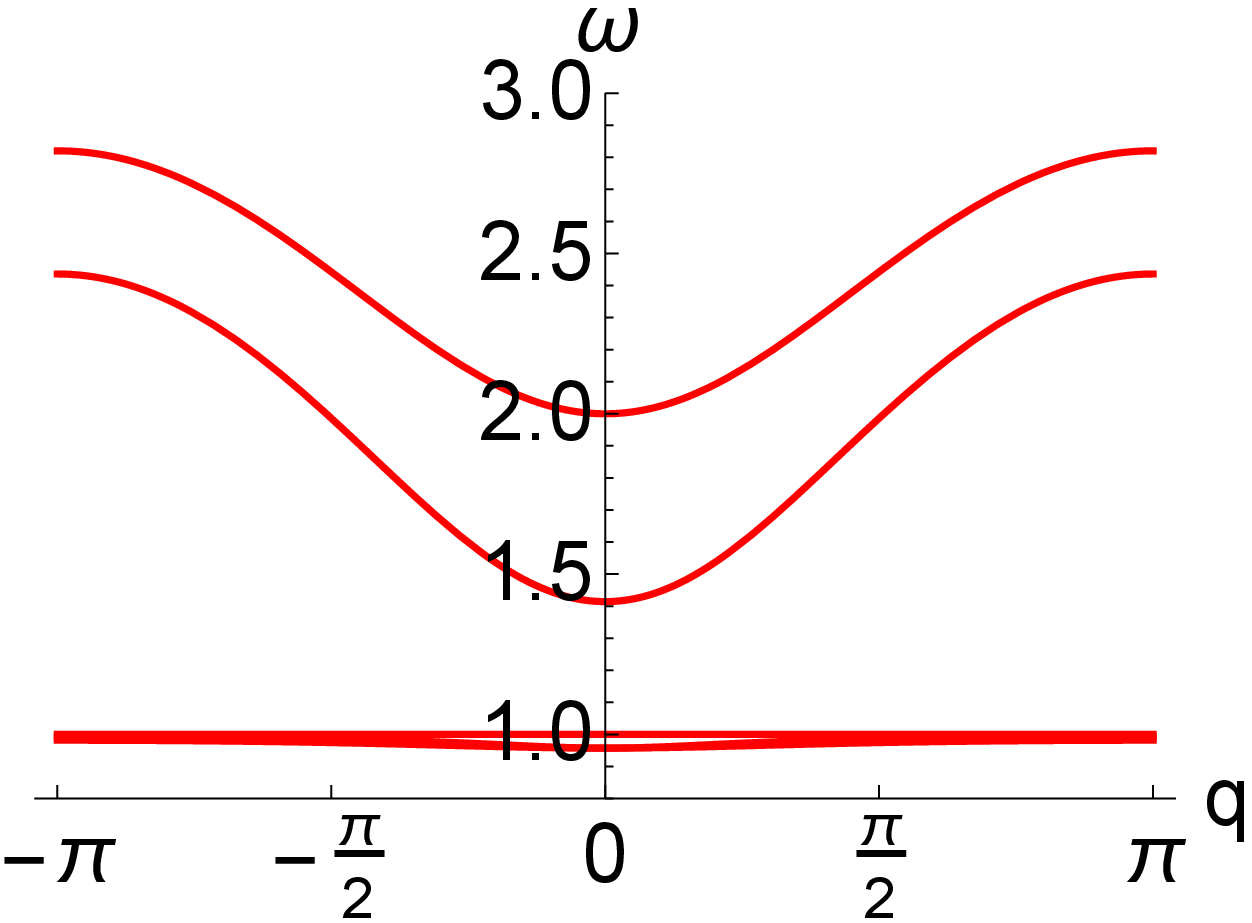}
		\caption{$ \beta_L = 1 $}
		\label{fig3b}
	\end{subfigure}
	\hskip1pt
	\begin{subfigure}{.32\textwidth}
		\centering
		\includegraphics[width=.99\linewidth]{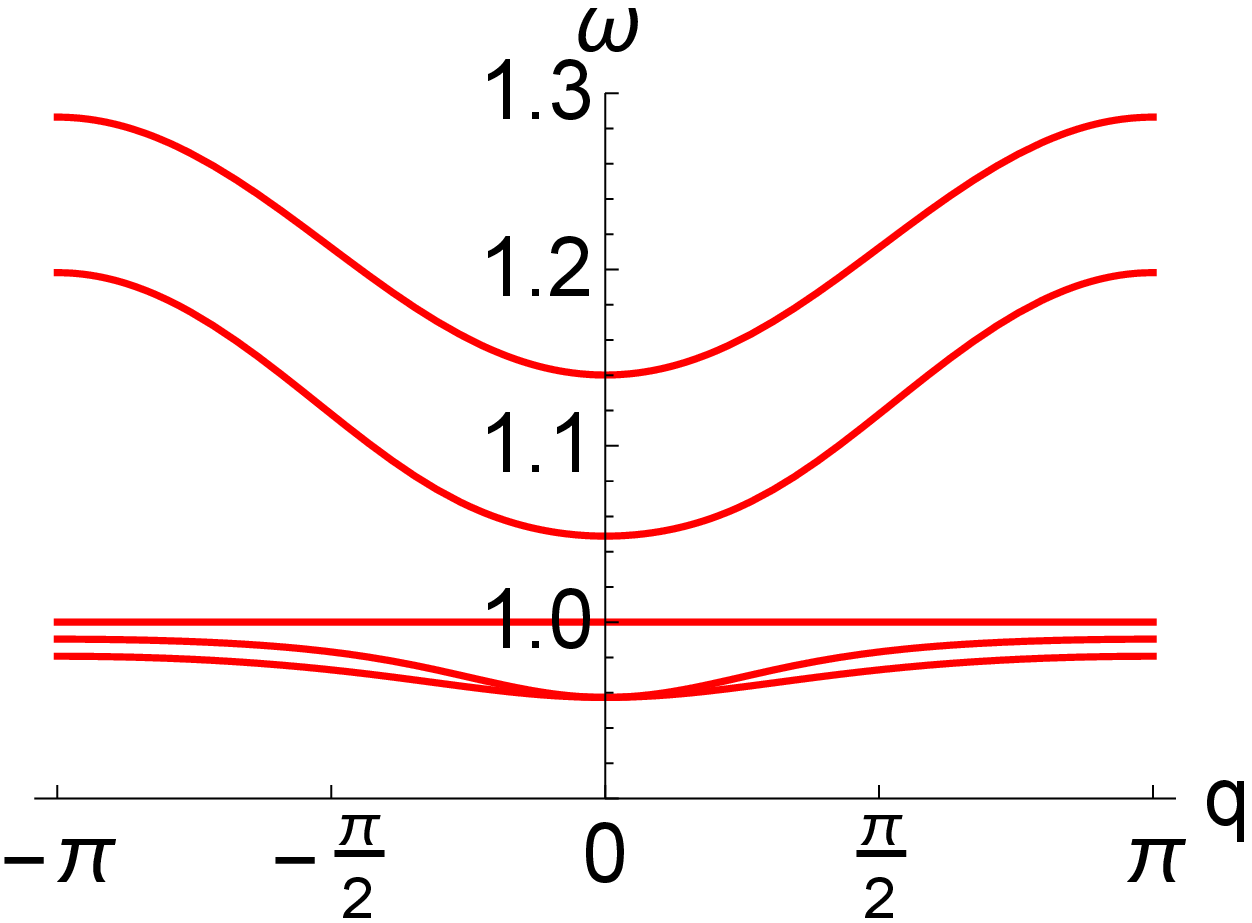}
		\caption{$ \beta_L = 10 $}
		\label{fig3c}
 \end{subfigure}%
\caption{Dispersion relation for the 3-row JJA at fixed values 
of the dc bias ($ \gamma=0.4$) and anisotropy ($~\eta=1$) and for the
 different values of the dimensionless inductance $ \beta_L $.}
\label{fig3}
\end{figure}

Next we discuss the properties of the plasmon spectrum as a function
of the discreteness parameter $\beta_L$. The respective plots for the 
different values of $\beta_L$ are given in figures \ref{fig3a}-\ref{fig3c}. 
Naturally, the increase of $\beta_L$ means larger
discreteness effects thus the absolute values of the plasmon
frequencies (for modes with $\omega_{n>0}$) increase greatly when $\beta_L$ 
decreases [see figure \ref{fig3a}]. In the opposite
limit $\beta_L \to \infty$ the interaction between the
cells dies out, hence the dispersion laws that lay above the 
$\omega_0=1$ branch flatten as can be seen
in figure \ref{fig3c}.
%@@@@@@@@@@@@@@@@@@@@@@@@@@@@@@@@@@@@@@@@@@@@@@@@@@@@@@@@@@@@@@@@@@@@
%
% Fig. 3 (fig3a-fig3c)
%
%____________________________________________________________________
%@@@@@@@@@@@@@@@@@@@@@@@@@@@@@@@@@@@@@@@@@@@@@@@@@@@@@@@@@@@@@@@@@@@@
On the other hand, the properties of the lower (almost flat) bands
depend mostly on the value of the dc bias as these branches
must satisfy $(1-\gamma^2)^{1/4}\le \omega_n<1$ for all $n<0$.

\begin{figure}[htb]
	\centering
	\begin{subfigure}{.32\textwidth}
		\centering
		\includegraphics[width=.99\linewidth]{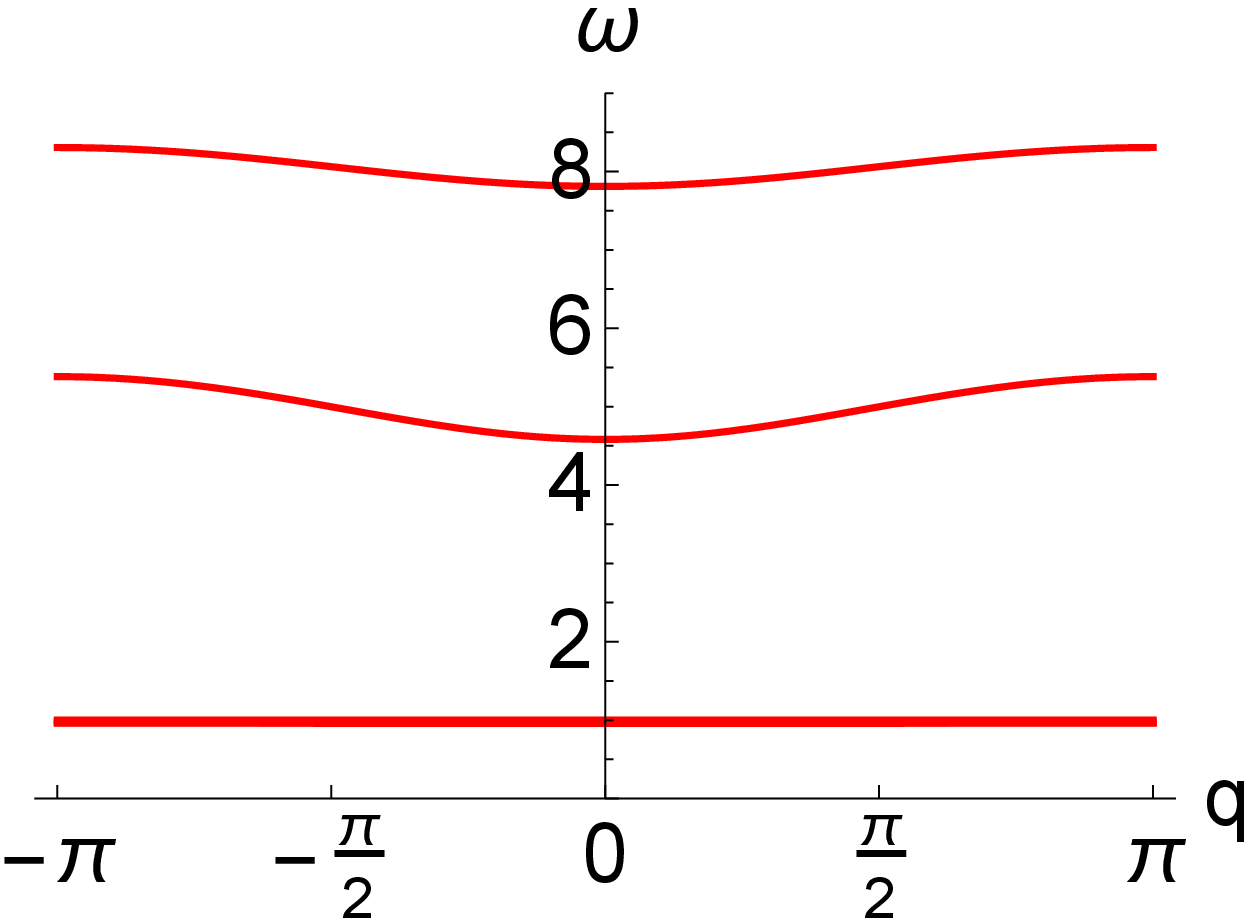}
		\caption{$\eta = 0.1$}
		\label{fig4a}
	\end{subfigure}
	\hskip1pt
%	\begin{subfigure}{.32\textwidth}
%		\centering
%		\includegraphics[width=.99\linewidth]{fig6b.eps}
%		\caption{$\beta_L = 1$}
%		\label{fig6b}
%	\end{subfigure}
	\hskip1pt
	\begin{subfigure}{.32\textwidth}
		\centering
		\includegraphics[width=.99\linewidth]{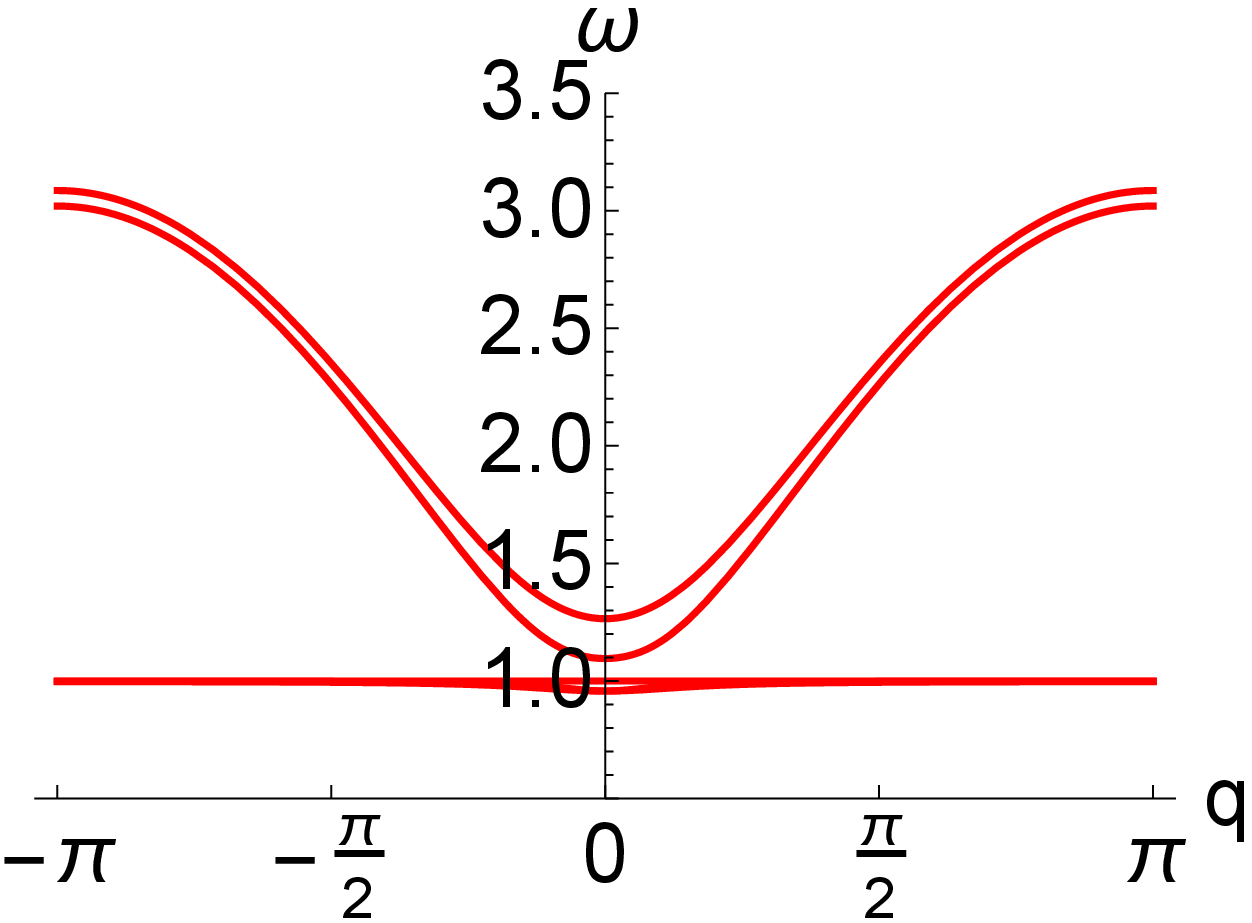}
		\caption{$\eta = 10$}
		\label{fig4b}
		\end{subfigure}
		
%		\centering
	\begin{subfigure}{.32\textwidth}
		\centering
		\includegraphics[width=.99\linewidth]{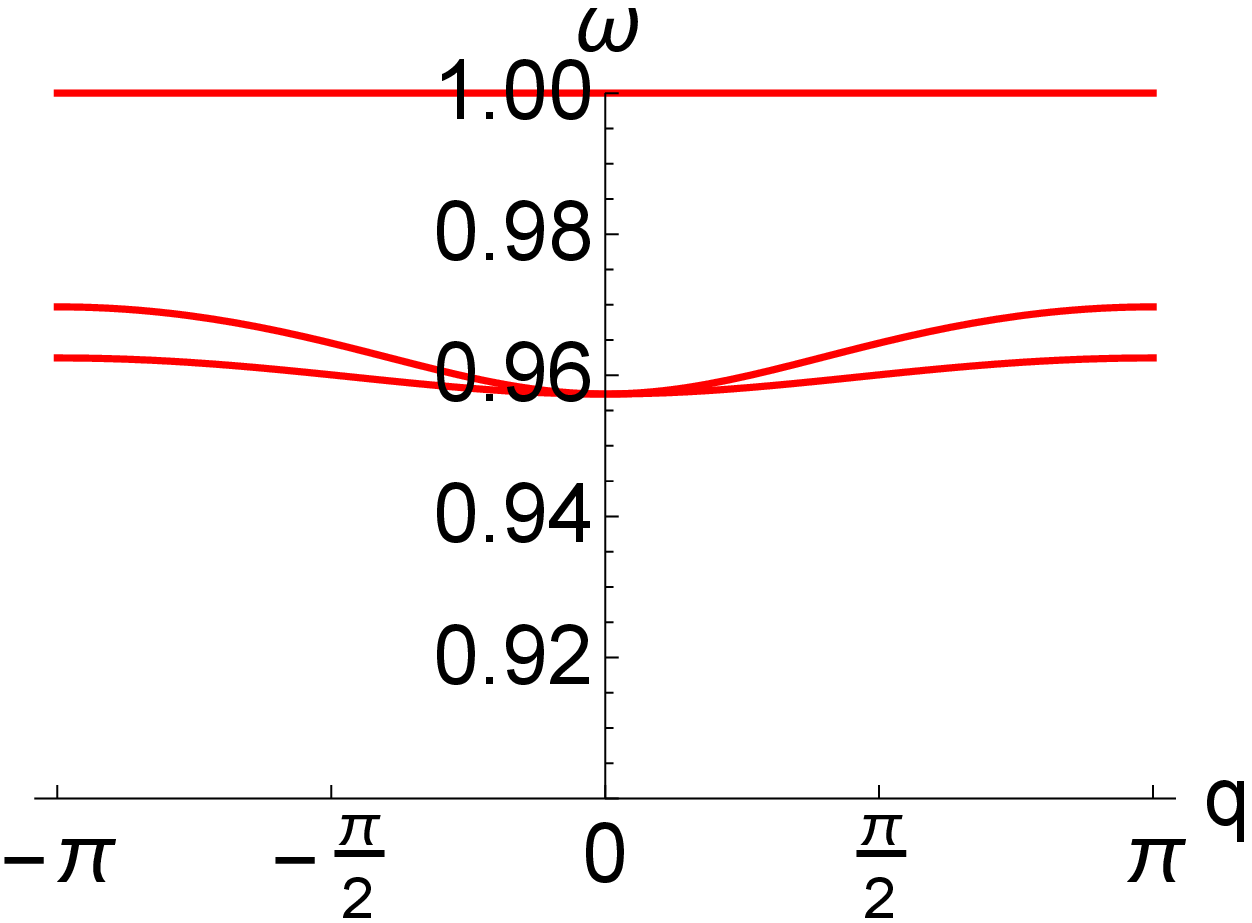}
		\caption{$\eta = 0.1$}
		\label{fig4c}
	\end{subfigure}
	\hskip1pt
	\begin{subfigure}{.32\textwidth}
		\centering
		\includegraphics[width=.99\linewidth]{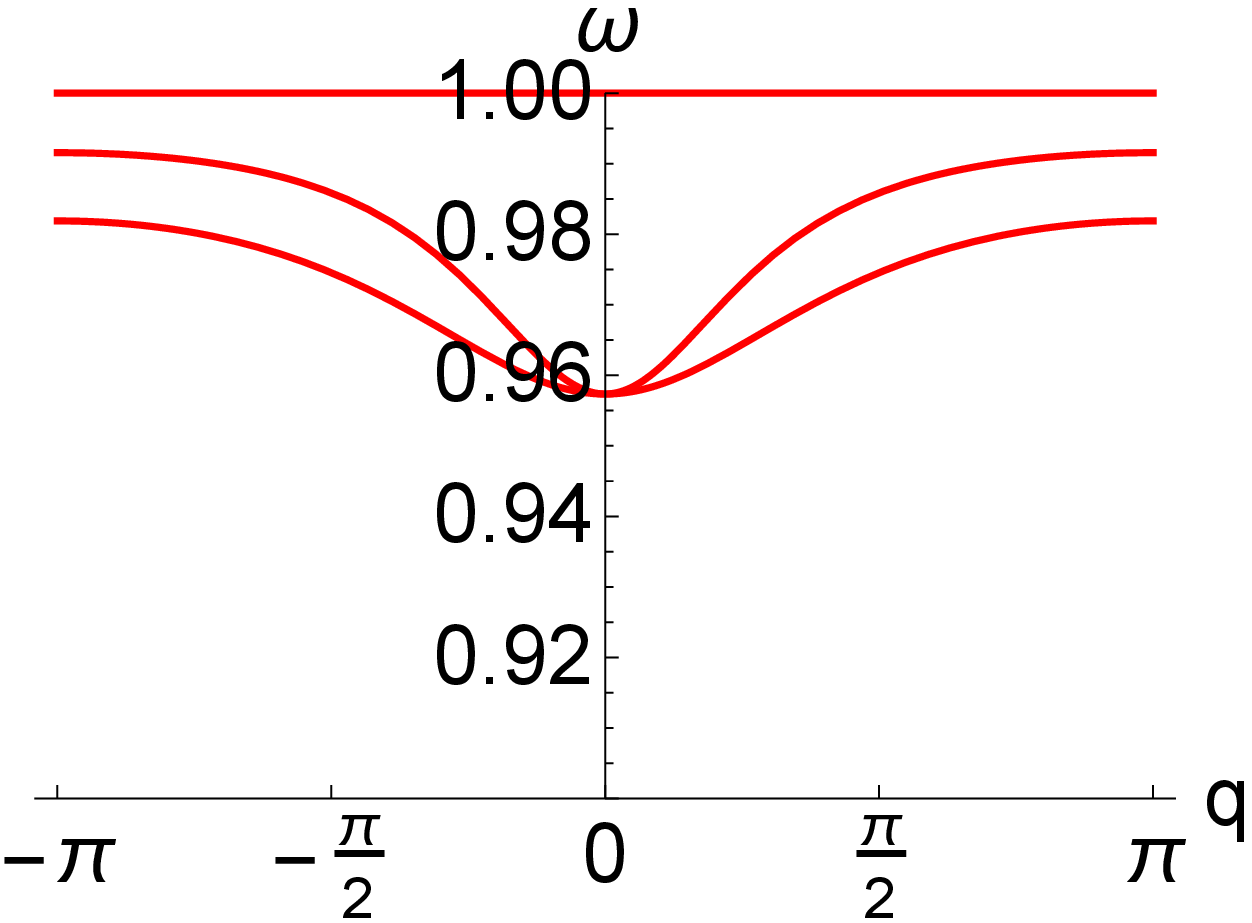}
		\caption{$\eta = 1$}
		\label{fig4d}
	\end{subfigure}
	\hskip1pt
	\begin{subfigure}{.32\textwidth}
		\centering
		\includegraphics[width=.99\linewidth]{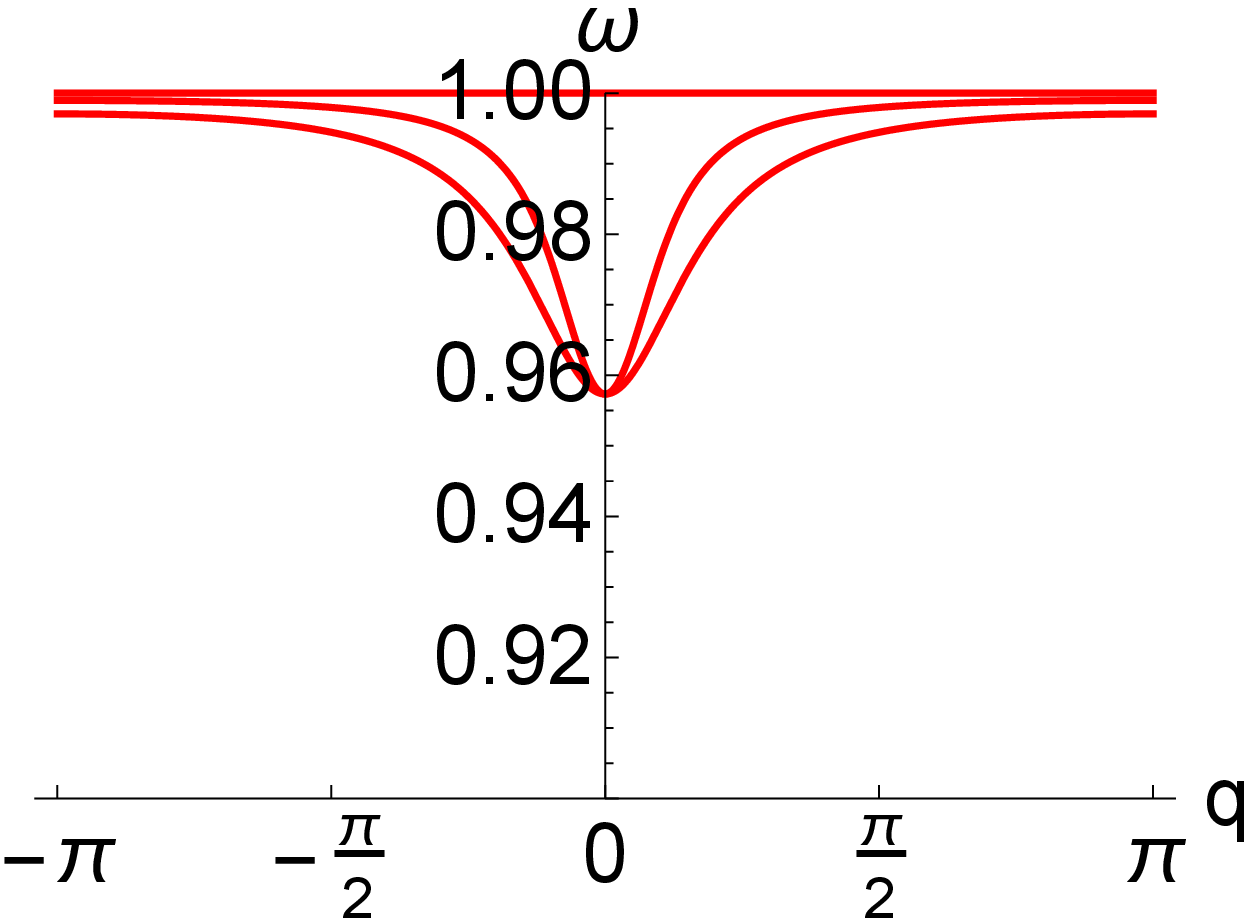}
		\caption{$\eta = 10$}
		\label{fig4e}
 \end{subfigure}
\caption{Dispersion relation for the 3-row JJA at the fixed values 
of the bias and inductance ($\gamma=0.4,\beta_L=0.5$) and 
for the different 
values of the anisotropy $\eta$. Figures (c)-(e) illustrate the
detailed view for the respective cases (see the subcations).}
\label{fig4}
\end{figure}

The role of the anisotropy constant $\eta$ is demonstrated in figures
\ref{fig4a}-\ref{fig4e}. This parameter controls the 
redistribution of the phase oscillations between the 
horizontal and vertical subsystems. 
%@@@@@@@@@@@@@@@@@@@@@@@@@@@@@@@@@@@@@@@@@@@@@@@@@@@@@@@@@@@@@@@@@@@@
%
% Fig. 4 (fig4a-fig4e)
%
%____________________________________________________________________
%@@@@@@@@@@@@@@@@@@@@@@@@@@@@@@@@@@@@@@@@@@@@@@@@@@@@@@@@@@@@@@@@@@@@
In the limit $\eta\to 0$ the horizontal oscillations dominate over the
vertical ones. First of all, the absolute values of the 
plasmon frequencies for the branches with $\omega_{n>0}>1$
increase significantly due to the presence of the $1/\eta$
singularity in the dispersion law. Moreover, the different branches
strongly separate from each other (see figure \ref{fig4a}). 
It is interesting that in the limit of small $\eta$ and intermediate
$\beta_L$ the dispersive branches (those that lie above $\omega_0=1$) 
have significant gaps between them.
The value of the gap between the $\omega_n$ and $\omega_{n+1}$
branches equals
%--------------------------------------------------------------------
\begin{eqnarray}\nonumber
&\Delta \omega_n=\omega_{n+1}(0)-\omega_n(\pi)\simeq
\sqrt{1+\frac{1+\alpha_{n+2}}{\eta\beta_L}}-\\
&-\sqrt{\frac{1+\alpha_{n+1}}{\eta\beta_L}} +{\cal O}(\eta^{1/2}).
\end{eqnarray}
%--------------------------------------------------------------------
This means that we have very weakly interacting horizontal
rows of junctions and the dispersion laws for them are strongly
separated from each other. In the opposite limit $\eta\to \infty$
the junctions in the vertical subsystem are coupled much stronger
as compared to the horizontal direction. Thus, there is very small
difference between the oscillations within one vertical 
column of junctions. As a result, this can be viewed as 
each column oscillating as a whole, and, consequently 
the branches $\omega_{n>0}$ in figure \ref{fig4b}
are very weakly separated and look almost identical.  

Finally we discuss the behavior of the almost flat
bands for the different values of anisotropy.
This behavior is illustrated in figures \ref{fig4c}-\ref{fig4e}.
The total width of the almost flat band ($\omega_{n<0}$) 
is defined as a difference between
the highest frequency value of the highest lying branch 
which is $\omega_{-1}(q)$ 
and the smallest value of the lowest branch
%--------------------------------------------------------------------
\begin{equation}
\Delta_-=\max_{q \in [-\pi,\pi]}\omega_{-1}(q)-\min_{q \in [-\pi,\pi]}\omega_{-N+1}(q)=\omega_{-1}(\pi)-(1-\gamma^2)^{1/4}.
\end{equation}
%--------------------------------------------------------------------
In the limit of small $\eta$ or small $\beta_L$ one can obtain the following
asymptotic behavior of the plasmon frequency at the Brillouin
zone edge:
%--------------------------------------------------------------------
\begin{equation}\label{20}
\omega_{n<0}(\pi)=\left \{\begin{array}{cc}
\left[1-\frac{1-\sqrt{1-\gamma^2}}{4\eta+1+\alpha_{n+1}}
(1+\alpha_{n+1}) \right ]^{1/2},& \beta_L\to 0,\\
(1-\gamma^2)^{1/4}, & \eta \to 0.
\end{array} \right.
\end{equation}
%--------------------------------------------------------------------
The limit $\eta\to 0$ is illustrated in figure \ref{fig4c}. In this
limit $\Delta_- \to 0$, hence, the whole band becomes flat.
If we keep the anisotropy $\eta$ constant and decrease the 
inductance the band width will depend on the number of rows, $N$, 
because $\alpha_2=1-2\cos{\pi/N}$. As a result, $\Delta_-$ does not
tend to $0$.

In the limit of large $\beta_L$ or $\eta$ the limiting
frequency value is given by
%--------------------------------------------------------------------
\begin{equation}\label{21}
\omega_{n<0}(\pi)=\left \{\begin{array}{c}
(1-\gamma^2)^{1/4}, \;\;\;\;\; \beta_L\to \infty,\\
 \left \{ \begin{array}{c}
1,\;\; \sqrt{1-\gamma^2}+\frac{4}{\beta_L}>1\\
\left[\sqrt{1-\gamma^2}+\frac{4}{\beta_L} \right ]^{1/2},
\sqrt{1-\gamma^2}+\frac{4}{\beta_L}<1
\end{array}\right.,\eta \to \infty.
\end{array} \right.
\end{equation}
%--------------------------------------------------------------------
In the limit when 
$\beta_L\to\infty$ for $\eta$ fixed we again expect complete
flattening of the $\omega_{n<0}$ band. 
The situation becomes more subtle if $\beta_L$ is finite
and $\eta \to \infty$. If discreteness is quite significant,
or, alternatively, the bias is strong enough to guarantee
the inequality $\beta_L>4/(1-\sqrt{1-\gamma^2})$ one
can obtain flattening when the strong discreteness limit
is taken:
\begin{equation}
\Delta_-=
\left[\sqrt{1-\gamma^2}+\frac{4}{\beta_L} \right ]^{1/2}-(1-\gamma^2)^{1/4} \to_{\beta_L\to \infty} 0.
\end{equation}
In the opposite case the band does not flatten because
$\Delta_-=1-(1-\gamma^2)^{1/4}$. This particular 
case is illustrated in figure \ref{fig4e} where the plasmon
branches have relatively sharp minimum for $|q|<1$ and
flatten only when $q$ approaches the Brillouin zone edge.

%####################################################################
\subsection{Spatial distribution of the Josephson phase vibrations}
\begin{figure}[htb]
 \begin{subfigure}{.19\textwidth}
  \setlength{\unitlength}{10.5cm}
  \begin{picture}(1.,0.43)
  \linethickness{0.3pt}
%\thicklines
%\matrixput(0.2, 0.0){4}(0.2, 0.15){2}{\circle*{0.025}}
  \multiput(0.03, 0.1)(0.16, 0.0){2}{\circle*{0.025}}
  \multiput(0.03, 0.25)(0.16, 0.0){2}{\circle*{0.025}}
  \multiput(0.03, 0.4)(0.16, 0.0){2}{\circle*{0.025}}

  \multiput(0.03,0.10)(0.16,0){2}{\line(0,1){0.3}}
  \multiput(0.0,0.1)(0,0.15){3}{\line(1,0){0.21}}

  \multiput(0.03, 0.175)(0.16,0){2}{\makebox(0,0){$\times$}}
  \multiput(0.03, 0.325)(0.16,0){2}{\makebox(0,0){$\times$}}
  \multiput(0.11, 0.1)(0.,0.15){3}{\makebox(0,0){$\times$}}	
  \linethickness{1.5pt}
  \color{red}{\multiput(0.05,0.1)(0,0.15){3}{\vector(1,0){0.09}}}
  \end{picture}
  \caption{$\omega_0$}
  \label{fig5a}  
 \end{subfigure}
 \begin{subfigure}{.19\textwidth}
  \setlength{\unitlength}{10.5cm}
  \begin{picture}(0.75,0.43)
%\thicklines
%\matrixput(0.2, 0.0){4}(0.2, 0.15){2}{\circle*{0.025}}
 \multiput(0.03, 0.1)(0.16, 0.0){2}{\circle*{0.025}}
 \multiput(0.03, 0.25)(0.16, 0.0){2}{\circle*{0.025}}
 \multiput(0.03, 0.4)(0.16, 0.0){2}{\circle*{0.025}}

 \multiput(0.03,0.10)(0.16,0){2}{\line(0,1){0.3}}
 \multiput(0.0,0.1)(0,0.15){3}{\line(1,0){0.21}}

 \multiput(0.03, 0.175)(0.16,0){2}{\makebox(0,0){$\times$}}
 \multiput(0.03, 0.325)(0.16,0){2}{\makebox(0,0){$\times$}}
 \multiput(0.11, 0.1)(0.,0.15){3}{\makebox(0,0){$\times$}}	
 \linethickness{1.5pt}
 \color{red}{\multiput(0.02,0.13)(0.16,0){2}{\vector(0,1){0.09}}}  
 \end{picture} 
 \caption{$\omega_{-2}$}
  \label{fig5b}  
 \end{subfigure} 
 \begin{subfigure}{.19\textwidth}
  \setlength{\unitlength}{10.5cm}
  \begin{picture}(0.5,0.43)
%\thicklines
%\matrixput(0.2, 0.0){4}(0.2, 0.15){2}{\circle*{0.025}}
 \multiput(0.03, 0.1)(0.16, 0.0){2}{\circle*{0.025}}
 \multiput(0.03, 0.25)(0.16, 0.0){2}{\circle*{0.025}}
 \multiput(0.03, 0.4)(0.16, 0.0){2}{\circle*{0.025}}

 \multiput(0.03,0.10)(0.16,0){2}{\line(0,1){0.3}}
 \multiput(0.0,0.1)(0,0.15){3}{\line(1,0){0.21}}

 \multiput(0.03, 0.175)(0.16,0){2}{\makebox(0,0){$\times$}}
 \multiput(0.03, 0.325)(0.16,0){2}{\makebox(0,0){$\times$}}
 \multiput(0.11, 0.1)(0.,0.15){3}{\makebox(0,0){$\times$}}	
 \linethickness{1.5pt}
% \color{red}{\put(0.02,0.25){\vector(0,1){0.07}}}  
 \color{red}{\multiput(0.02,0.28)(0.16,0){2}{\vector(0,1){0.09}}}  
 \end{picture}
  \caption{$\omega_{-1}$}
  \label{fig5c}
 \end{subfigure} 
%
% d
%
 \begin{subfigure}{.19\textwidth}
  \setlength{\unitlength}{10.5cm}
  \begin{picture}(0.75,0.43)
%\thicklines
%\matrixput(0.2, 0.0){4}(0.2, 0.15){2}{\circle*{0.025}}
\multiput(0.03, 0.1)(0.16, 0.0){2}{\circle*{0.025}}
\multiput(0.03, 0.25)(0.16, 0.0){2}{\circle*{0.025}}
\multiput(0.03, 0.4)(0.16, 0.0){2}{\circle*{0.025}}

\multiput(0.03,0.10)(0.16,0){2}{\line(0,1){0.3}}
\multiput(0.0,0.1)(0,0.15){3}{\line(1,0){0.21}}

  \multiput(0.03, 0.175)(0.16,0){2}{\makebox(0,0){$\times$}}
  \multiput(0.03, 0.325)(0.16,0){2}{\makebox(0,0){$\times$}}
  \multiput(0.11, 0.1)(0.,0.15){3}{\makebox(0,0){$\times$}}	
  \linethickness{1.5pt}  
  \color{red}{\put(0.05,0.1){\vector(1,0){0.09}}}
  \color{blue}{\put(0.14,0.4){\vector(-1,0){0.09}}}  
  \end{picture}
   \caption{$\omega_{+1}$}
  \label{fig5d}
  \end{subfigure} 
 \begin{subfigure}{.19\textwidth}
  \setlength{\unitlength}{10.5cm}
  \begin{picture}(0.5,0.43)
%\thicklines
%\matrixput(0.2, 0.0){4}(0.2, 0.15){2}{\circle*{0.025}}
 \multiput(0.03, 0.1)(0.16, 0.0){2}{\circle*{0.025}}
 \multiput(0.03, 0.25)(0.16, 0.0){2}{\circle*{0.025}}
 \multiput(0.03, 0.4)(0.16, 0.0){2}{\circle*{0.025}}

 \multiput(0.03,0.10)(0.16,0){2}{\line(0,1){0.3}}
 \multiput(0.0,0.1)(0,0.15){3}{\line(1,0){0.21}}

 \multiput(0.03, 0.175)(0.16,0){2}{\makebox(0,0){$\times$}}
 \multiput(0.03, 0.325)(0.16,0){2}{\makebox(0,0){$\times$}}
 \multiput(0.11, 0.1)(0.,0.15){3}{\makebox(0,0){$\times$}}	
 \linethickness{1.5pt}  
  \color{red}{\multiput(0.05,0.1)(0,0.3){2}{\vector(1,0){0.05}}}
  \color{blue}{\put(0.15,0.25){\vector(-1,0){0.09}}}
%  \color{red}{\put(0.0,0.4){\vector(1,0){0.07}}} 
 \end{picture}
 \caption{$\omega_{+2}$}
  \label{fig5e} 
 \end{subfigure} 
\caption{Schematic representation of the plasmon amplitude
distribution for the array with $N=3$ rows in the long wave
limit $q=0$. Color arrows correspond
to the excited junctions where the
red arrow (pointing right) and blue arrow (pointing left) 
represent positive (negative) 
amplitude of the respective junction.
Arrows directed upwards describe excited vertical junctions.
Unexcited junctions have no arrows. Subcaptions under each figure
denote to which eigenvalue this eigenvector belongs.}
\label{fig5} 
\end{figure}
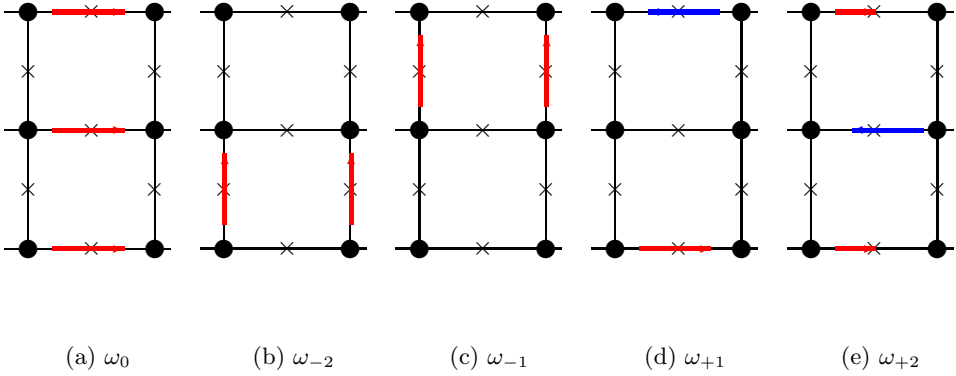

%####################################################################
The respective eigenvectors $\left(\mathbf{A}^{(v)},
\mathbf{A}^{(h)}\right)=\left(A^{(v)}_1, \cdots,  A^{(v)}_{N-1}, 
A^{(h)}_1, \cdots, A^{(h)}_{N}\right)$ 
can be computed from the linear set
of equations that emanates from the equations of motion 
(\ref{8})-(\ref{11}). With the help of the matrices
$L$, $U$ and $D_{v,h}$ defined in \ref{app-a} it is possible to write
the equations for the eigenvectors in the
following form:
%--------------------------23-26-------------------------------------
\begin{eqnarray}\label{23}
&& [\omega_i^2(q) - \omega^2]A_n^{(v)}+\frac{1-e^{-iq}}{\beta_L} 
 \left( A_n^{(h)}-A_{n+1}^{(h)}\right)=0,\; n=\overline{1,N-1},\\
\label{24}
&& ({1-e^{iq}}) \left (A_n^{(v)}-A_{n-1}^{(v)}\right )- A_{n-1}^{(h)}- A^{(h)}_{n+1}+\\
\nonumber 
&&+[2\beta_L\eta(1-\omega^2)+1]A_n^{(h)}=0,\;n=\overline{2,N-1},\\
\label{25}
&&({1-e^{iq}}) A_1^{(v)}+ [\beta_L\eta(1-\omega^2)+1]
A_1^{(h)}-A_2^{(h)}=0,\\
\label{26}
&&({1-e^{iq}}) A_N^{(v)} +A_{N-1}^{(h)}-[\beta_L\eta(1-\omega^2)+1]
A_N^{(h)}=0 ~.
\end{eqnarray}
%--------------------------------------------------------------------
\paragraph{Long-wave limit $q=0$.} In this limit regardless of the
particular eigenfrequency the plasmon modes of the vertical and 
horizontal sublattices decouple. As a result, the amplitudes
of the vertical subsystem come as a solution of the strongly
degenerate system of equations and are basically the eigenvectors
of the diagonal matrix:
%[in the case when $\omega\neq \sqrt{1-\gamma^2}$]: 
$\mathbf{A}^{(v)}=[(1,0,\dots,0)$,$(0,1,\dots,0)$,$\ldots$,
$(0,0,\dots,1)]$.
In the horizontal subsystem all the junctions are excited
and their amplitudes satisfy more complex conditions as they
are eigenvectors of the tridiagonal matrix.
One of the eigenfrequencies coincides with the Josephson
plasma frequency ($\omega_0=1$) and is dispersionless. The respective
eigenvector consists of only horizontal junctions exited
while all vertical ones are at rest as shown schematically in
figure \ref{fig5a}.  
%In the particular degenerate case of $\gamma=0$ we discuss the
%degenerate eigenfrequencies with $\omega_n=1$. 
In the most degenerate case when the external bias is absent
the general picture shown in figures \ref{fig5a}-\ref{fig5e} does not change.

%@@@@@@@@@@@@@@@@@@@@@@@@@@@@@@@@@@@@@@@@@@@@@@@@@@@@@@@@@@@@@@@@@@@@
%
%
%  Fig. 5 (fig 5a-fig5e)
%
%
%____________________________________________________________________
%@@@@@@@@@@@@@@@@@@@@@@@@@@@@@@@@@@@@@@@@@@@@@@@@@@@@@@@@@@@@@@@@@@@@

\paragraph*{Beyond the long-wave limit ($q\neq 0$).}
If one departs slightly from the center of the Brillouin zone ($q=0$)
the vibrations in the horizontal and vertical subsystems 
start to mix. For small values of dc bias the terms 
$\propto (1-\cos q)$ start to appear in the components that were
unexcited in the $q=0$ limit. 
In the unbiased case there is $(N-1)$-fold
degeneracy of the $\omega=1$ eigenfrequency.
The mode which initially 
was uniform with respect to horizontal vibrations 
and with no excited vertical vibrations now has both vertical and horizontal 
junctions excited. Similarly, the $\omega_0=1$ mode has also both 
vertical and horizontal junctions excited. 

\begin{figure}[htb]
 \begin{subfigure}{.32\textwidth}
   \setlength{\unitlength}{10.5cm}
   \begin{picture}(1.,0.43)
%\thicklines
%\matrixput(0.2, 0.0){4}(0.2, 0.15){2}{\circle*{0.025}}
  \multiput(0.1, 0.1)(0.2, 0.0){2}{\circle*{0.025}}
  \multiput(0.1, 0.25)(0.2, 0.0){2}{\circle*{0.025}}
  \multiput(0.1, 0.4)(0.2, 0.0){2}{\circle*{0.025}}

  \multiput(0.1,0.10)(0.2,0){2}{\line(0,1){0.3}}
  \multiput(0.05,0.1)(0,0.15){3}{\line(1,0){0.30}}

  \multiput(0.1, 0.175)(0.2,0){2}{\makebox(0,0){$\times$}}
  \multiput(0.1, 0.325)(0.2,0){2}{\makebox(0,0){$\times$}}
  \multiput(0.2, 0.1)(0.,0.15){3}{\makebox(0,0){$\times$}}	
   \linethickness{1.5pt}
  \color{red}{\multiput(0.1,0.1)(0,0.15){3}{\vector(1,0){0.1}}}
  \end{picture}
  \caption{$\omega_0=1$}
  \label{fig6a} 
  \end{subfigure}
 \begin{subfigure}{.32\textwidth}
   \setlength{\unitlength}{10.5cm}
   \begin{picture}(1.,0.43)
%\thicklines
%\matrixput(0.2, 0.0){4}(0.2, 0.15){2}{\circle*{0.025}}
   \multiput(0.1, 0.1)(0.2, 0.0){2}{\circle*{0.025}}
   \multiput(0.1, 0.25)(0.2, 0.0){2}{\circle*{0.025}}
   \multiput(0.1, 0.4)(0.2, 0.0){2}{\circle*{0.025}}

   \multiput(0.1,0.10)(0.2,0){2}{\line(0,1){0.3}}
   \multiput(0.05,0.1)(0,0.15){3}{\line(1,0){0.30}}

   \multiput(0.1, 0.175)(0.2,0){2}{\makebox(0,0){$\times$}}
   \multiput(0.1, 0.325)(0.2,0){2}{\makebox(0,0){$\times$}}
   \multiput(0.2, 0.1)(0.,0.15){3}{\makebox(0,0){$\times$}}	
   \linethickness{1.5pt}
   \color{red}{\multiput(0.09,0.1)(0,0.15){2}{\vector(0,1){0.09}}}
   \color{red}{\multiput(0.28,0.1)(0,0.15){2}{\vector(0,1){0.09}}}   
   \color{red}{\put(0.1,0.1){\vector(1,0){0.1}}}
   \color{blue}{\put(0.23,0.4){\vector(-1,0){0.1}}}
%   \psline[linecolor=red,linewidth=1.5pt]{->}(0.05cm,0.05cm)(2,0)
  \end{picture}
  \caption{$\omega_{\pm 1}$}
  \label{fig6b} 
 \end{subfigure}
 \begin{subfigure}{.32\textwidth}
  \setlength{\unitlength}{10.5cm}
   \begin{picture}(1.,0.43)
%\thicklines
%\matrixput(0.2, 0.0){4}(0.2, 0.15){2}{\circle*{0.025}}
   \multiput(0.1, 0.1)(0.2, 0.0){2}{\circle*{0.025}}
   \multiput(0.1, 0.25)(0.2, 0.0){2}{\circle*{0.025}}
   \multiput(0.1, 0.4)(0.2, 0.0){2}{\circle*{0.025}}

   \multiput(0.1,0.10)(0.2,0){2}{\line(0,1){0.3}}
   \multiput(0.05,0.1)(0,0.15){3}{\line(1,0){0.30}}

   \multiput(0.1, 0.175)(0.2,0){2}{\makebox(0,0){$\times$}}
   \multiput(0.1, 0.325)(0.2,0){2}{\makebox(0,0){$\times$}}
   \multiput(0.2, 0.1)(0.,0.15){3}{\makebox(0,0){$\times$}}	
   \linethickness{1.5pt}
   \color{red}{\multiput(0.09,0.1)(0.2,0.0){2}{\vector(0,1){0.09}}}
   \color{blue}{\multiput(0.08,0.4)(0.2,0.0){2}{\vector(0,-1){0.09}}}   
   \color{red}{\multiput(0.1,0.1)(0,0.3){2}{\vector(1,0){0.07}}}
   \color{blue}{\put(0.23,0.25){\vector(-1,0){0.12}}}
  \end{picture}
  \subcaption{$\omega_{\pm 2}$} 
  \label{fig6c}  
 \end{subfigure}
\caption{Schematic representation of the plasmon waves
that correspond to different eigenfrequencies (see the respective
subcaption) for $N=3$ rows and $\gamma\neq 0$. The meaning of
the color arrows is the same as in figures \ref{fig5a}-\ref{fig5e}.}
\end{figure}
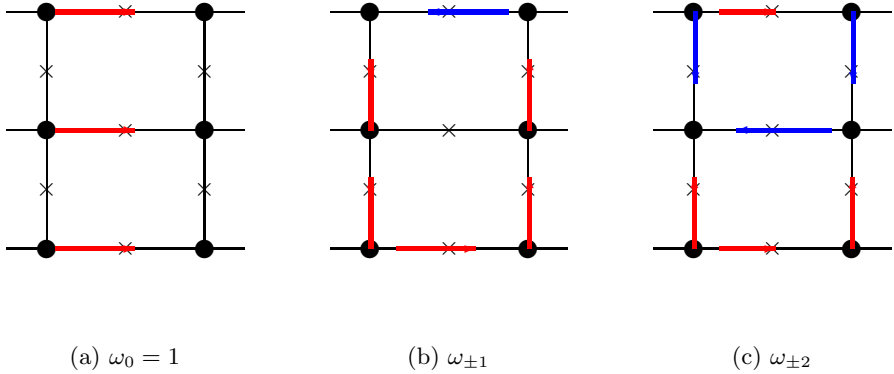

If the external bias $\gamma$ is applied the degeneracy for the modes
$\omega_{-N+1},\ldots,\omega_{-1}$
is lifted (except the central point $q=0$).
As a result, the $\omega_0=1$ mode retains the same structure
as it had in the long-wave limit: $A_n^{(v)}=0$, $n=\overline{1,N-1}$;
$A_n^{(h)}=A\neq 0$, $n=\overline{1,N}$ (see also figure \ref{fig5a}).
The modes that lie below the flat branch $\omega_{n}(q)<1, n<0$ have the same spatial 
structure as their counterparts above the flat branch. In other words,
the spatial structure of $\omega_{-n}$ and $\omega_{n}$ modes will be
the same for each $n$. The particular case for $N=3$ is shown in   
figures \ref{fig6a}-\ref{fig6c}.
The modes with $\omega_{\neq 0}$ have certain symmetry with 
respect to the symmetry line that is parallel to the $OX$ axis
and cuts the array in halves. If the number of rows $N$ is odd this line
coincides with the $(N+1)\,$th row while if $N$ is even this line
lies in the middle between the $N/2\,$th and $(N+1)/2\,$th rows.
According to this symmetry, in the amplitude distribution of each mode
either the vertical phases are excited in
the antisymmetric way while the horizontal phases are excited in 
the symmetric way or vice versa. The figures \ref{fig6b}-\ref{fig6c}
clearly demonstrate that.
%@@@@@@@@@@@@@@@@@@@@@@@@@@@@@@@@@@@@@@@@@@@@@@@@@@@@@@@@@@@@@@@@@@@@
%
%
% Fig. 6
%
%
%____________________________________________________________________
%@@@@@@@@@@@@@@@@@@@@@@@@@@@@@@@@@@@@@@@@@@@@@@@@@@@@@@@@@@@@@@@@@@@@
At the edges of the Brillouin zone ($q=\pm \pi$) the structure of the 
non-flat bands becomes more symmetric with all the amplitudes 
of the horizontal junctions having the same absolute value.

%####################################################################
%####################################################################
\section{Discussion and conclusions}
%####################################################################
%####################################################################

In this article the linear wave spectrum of the Q1D JJA is studied. The
Q1D JJA is a multi-ladder array that is considered to be infinite
in $X$ direction and consists of $N$ rows in $Y$ direction.
The array is uniformly biased by the dc current applied along
the every vertical column. The main result can be summarized 
in the following way. The Josephson plasmon spectrum of the
array consists of $2N-1$ branches. In the unbiased case
$N$ branches are completely flat and the respective eigenfrequency
coincides with the Josephson plasma frequency. The remaining
$N-1$ branches are strongly dispersive and their dispersion
laws are similar to the standard 1D JJA dispersion relations.
When the dc bias is applied the $N$-fold degeneracy is lifted
and only one flat branch remains. This flat branch corresponds to the
plasmon mode where all vertical junctions are not excited 
and all horizontal junctions oscillate in phase.
The rest of the branches
become weakly dispersive if the dc bias is small: $\gamma=I_B/I_c^{(v)}\ll 1$. 
In that case the width of all flat bands is confined
by the limits $(1-\gamma^2)^{1/4}\le \omega < 1$. 

The obtained Q1D JJA spectra are important for the studies of
discrete breathers in these structures. While the discrete breather properties
in the simple JJ ladder are well understood, 
their counterparts in more complex ladder-like structures have not been
studied yet. The current studies of the linear spectra are the necessary
first step in that direction. In particular, we would like to
point out that in the limit of small horizontal-vertical anisotropy $\eta\ll 1$
there are large gaps between the plasmon modes. This opens a possibility
of the gap discrete breathers \cite{kf92pra} appearing in different gaps
of the linear spectrum. Finally, the phenomenon of Fano resonances
in the simlest JJ ladder \cite{msffu05prb} can be investigated for more
complex Q1D JJAs.

\ack
Both the authors acknowledge the support by the National
Research Foundation of Ukraine grant (2020.02/0051)
"Topological phases of matter and excitations in Dirac
materials, Josephson junctions and magnets".

%####################################################################
\appendix
\section{Computation of the characteristic polynomial}
\label{app-a}
%####################################################################

This Appendix is devoted to the details of the characteristic polynomial
computation. The characteristic polynomial is given by the
determinant of the $2N-1 \times 2N-1$ matrix
%--------------------------------------------------------------------
\begin{equation} \label{A1}
\chi(\omega^2)=\det (A-\mathbb{I}_{2N-1}\omega^2),\;
A= \left[ 
\begin{array}{c|c}
D_v  & U \\ \hline
L    & D_h 
\end{array}\right].
\end{equation}
%--------------------------------------------------------------------
Here $D_v=\omega^2_i(q) \mathbb{I}_{N-1}$, 
$\mathbb{I}_N$ is the $N\times N$ identity matrix, $U$ is $N\times N-1$
matrix, $L$ is $N-1\times N$ matrix and $D_h$ is tridiagonal $N-1\times N-1$. 
These three matrices are given by the following expressions:
%--------------------------------------------------------------------
\begin{eqnarray}
&&{D_h}\equiv\left[\begin{array}{ccccccc}\vspace{2mm}
1+b&-b&0&\cdots &0&0&0\\ \vspace{2mm}
-b&1+2 b&-b&\cdots &0&0&0\\ \vspace{2mm}
0&-b&1+2 b&\cdots &0&0&0\\ \vspace{2mm}
\vdots &\vdots &\vdots &\ddots & \vdots&\vdots&\vdots\\ \vspace{2mm}
0&0&0&\cdots &-b&1+2 b&-b\\ \vspace{2mm}
0&0&0&\cdots &0&-b&1+b\\
\end{array}\right], \\
&& U\equiv \frac{1-e^{-i q}}{\beta_L}~ S, \;S= \left[\begin{array}{ccccccc} \vspace{2mm}
1&-1&0 &\cdots &0&0&0\\ \vspace{2mm}
0& 1&-1&\cdots &0&0&0\\ \vspace{2mm}
0& 0&1 &\cdots &0&0&0\\ \vspace{2mm}
\vdots &\vdots &\vdots &\ddots & \vdots&\vdots &\vdots\\ \vspace{2mm}
0& 0& 0&\cdots &1 &-1&0\\ \vspace{2mm}
0& 0& 0&\cdots &0 & 1&-1\\ 
\end{array}\right], \\
&& L\equiv \frac{1-e^{i q}}{\beta_L \eta}~ S^T,\;b \equiv \frac{1}{\beta_L  \eta }.
\end{eqnarray}
%--------------------------------------------------------------------
The term $\omega_i(q)$ is the dispersion law of the one-dimensional
array of the biased vertical Josephson junctions (see \cite{u98pd,wszo95prl})
%--------------------------------------------------------------------
\begin{equation}\label{JJ:18}
\omega^2_i(q) \equiv \sqrt{1-\gamma ^2}+\frac{2}{\beta_L } (1-\cos q).
\end{equation}
%--------------------------------------------------------------------
With the help of Schur complement \cite{zhang05} it is possible to 
simplify the determinant of the matrix $A-\mathbb{I}_{2N-1}\omega^2$
[see equation (\ref{A1})]:
%--------------------------------------------------------------------
\begin{eqnarray}\nonumber
&&\chi(\omega^2)=\det (D_v-\omega^2\mathbb{I}_{N-1} )\det [D_h-\omega^2\mathbb{I}_N-L (D_v-\omega^2\mathbb{I}_{N-1})^{-1} U] =\\
&&=[\omega_i^2(q)-\omega^2]^{N-1} \det \left [D_h-\omega^2\mathbb{I}_N- 
\frac{1}{(\omega_i^2(q)-\omega^2)}LU \right ],\\
&&LU=\frac{|1-e^{iq}|^2}{\eta \beta_L^2}\left[\begin{array}{ccccccc} \vspace{2mm}
1&-1&0 &\cdots &0&0&0\\ \vspace{2mm}
-1& 2&-1&\cdots &0&0&0\\ \vspace{2mm}
0& -1&2 &\cdots &0&0&0\\ \vspace{2mm}
\vdots &\vdots &\vdots &\ddots & \vdots&\vdots &\vdots\\ \vspace{2mm}
0& 0& 0&\cdots &-1 &2&-1\\ \vspace{2mm}
0& 0& 0&\cdots &0 & -1&1\\ 
\end{array}\right].
\end{eqnarray}
%--------------------------------------------------------------------
The matrix in $[\cdots]$ is an $N\times N$ tridiagonal matrix.
The resulting characteristic polynomial can be written as a
determinant of the tridiagonal matrix $D_N$:
%--------------------------------------------------------------------
\begin{eqnarray}
&&\chi(\omega^2)=\frac{\det{D_N}}{\eta^N [\omega_i(q)-\omega]},  \;
\\
%&&\det{D_N}=d_0 {f_{N-1}}-d_1^2 {f_{N-2}}, N=3,4,\ldots. ,\\
%&&f_n=d_2 f_{n-1}-d_1^2f_{n-2}, n=2,3,\ldots,N-1,\; f_0=1,f_1=d_0,
&& D_2= \left [\begin{array}{cc}
d_0 & d_1 \\ d_1 & d_0
\end{array} \right ],\;
D_N=\left[
\begin{array}{ccccccc}
d_0 & d_1 &  0  &  \cdots & 0 & 0 & 0 \\
d_1 & d_2 & d_1 &  \cdots & 0 & 0 & 0 \\
0   & d_1 & d_2 &  \cdots & 0 & 0 & 0  \\  
\vdots  &\vdots   &\vdots  & \ddots & \vdots & \vdots & \vdots \\
0   & 0   & 0 & \cdots & d_2 & d_1 & 0    \\ 
0   & 0   & 0 & \cdots & d_1 & d_2 & d_1  \\ 
0   & 0   & 0 & \cdots & 0   & d_1 & d_0  \\
\end{array} \right],
\end{eqnarray}
%--------------------------------------------------------------------
with
%--------------------------------------------------------------------
\begin{eqnarray}
\label{A10}
&&d_0=\eta [\omega^2_i(q)-\omega^2] \left (1+\frac{1}{\eta\beta_L}-\omega^2\right )-\frac{2(1-\cos{q})}{\beta_L^2},  \\
\label{A11}
&&d_1=- \frac{1}{\beta_L}  [\omega^2_i(q)-\omega^2]+\frac{2(1-\cos{q})}{\beta_L^2}, \\
&&d_2=d_0-d_1=\eta [\omega^2_i(q)-\omega^2] \left (
1+\frac{2}{\eta\beta_L}-\omega^2\right )-\frac{4(1-\cos{q})}{\beta_L^2}, 
\end{eqnarray}
%--------------------------------------------------------------------
Note that the matrix $D_2$ does not contain the $d_2$ element

In general, the determinant $\det D_N$ can be factorized if the
matrix $D_N$ can be diagonalized. The eigenvalues of this tridiagonal
matrix are known \cite{y05ame-notes}:
%--------------------------------------------------------------------
\begin{equation}
\lambda_n=d_0+\left[-1+2\cos  \frac{\pi (n-1)}{N}  \right ]d_1,\;
n=\overline{1,N}\,.
\end{equation}
%--------------------------------------------------------------------
Therefore, the determinant of the matrix $D_N$ can be written 
explicitly. Also it should be taken into accound that there
always exists the first eigenvalue, $\lambda_1=d_0+d_1$. This
will help to remove the singular term in the characteristic polynomial.
%--------------------------------------------------------------------
\begin{eqnarray}\nonumber
&&\det D_N=\prod_{n=1}^{N} \left \{d_0+\left[-1+\cos \frac{\pi (n-1)}{N} 
\right ]d_1\right \}=
(d_0+d_1) \times \\
&&\times \prod_{n=2}^{N} \left \{d_0+\left[-1+\cos\frac{\pi (n-1)}{N}\right ]d_1\right \}.
\end{eqnarray}
%--------------------------------------------------------------------
As a result, the characteristic polynomial for the squared frequency
can be written as a product:
%--------------------------------------------------------------------
\begin{eqnarray}
\chi(\omega^2)\propto (\omega^2-1)\prod_{n=2}^{N} [d_0-\alpha_n d_1],~~ \alpha_n=1-2\cos \frac{\pi (n-1)}{N}.
\end{eqnarray}
%--------------------------------------------------------------------
The roots of this polynomial are given by the equation $\omega^2=1$
and equation $d_0=\alpha_n d_1$. Note that $d_0$ and $d_1$ are
polynomials for $\omega^2$ given by equations (\ref{A10})-(\ref{A11}). 
Equation $d_0=\alpha_n d_1$ transforms 
(the subscript $_n$ has been dropped for the sake of simplicity)
into the quadratic equation for $\omega^2$: 
%--------------------------------------------------------------------
\begin{eqnarray}\nonumber
\omega^4&-&\left(1+\omega_i^2+\frac{1+\alpha}{\eta\beta_L} \right)\omega^2
+\omega_i^2 \left(1+\frac{1+\alpha}{\eta\beta_L} \right)-\\
&-&2(1+\alpha)\frac{1-\cos{q}}{\eta\beta_L^2}=0.
\end{eqnarray}
%--------------------------------------------------------------------
The roots of the abovementioned equation  are
%--------------------------------------------------------------------
\begin{eqnarray}\nonumber
\omega^2(q)&=&\frac{1}{2}\left(1+\omega_i^2+\frac{1+\alpha}{\eta\beta_L} \right) \pm \\ 
&\pm & \sqrt{\frac{1}{4}\left[\omega_i^2-\left(1+\frac{1+\alpha}{\eta\beta_L} \right)\right ]^2+2(1+\alpha)\frac{1-\cos q}{\eta\beta_L^2}}.
\end{eqnarray}
%--------------------------------------------------------------------

\section*{ORCID iDs}
Daryna Bukatova \href{https://orcid.org/0000-0002-0522-851X}{https://orcid.org/0000-0002-0522-851X}

\noindent
Yaroslav Zolotaryuk \href{https://orcid.org/0000-0003-1079-0221}{https://orcid.org/0000-0003-1079-0221}

\section*{References}
%\References
%

%\bibliographystyle{unsrt}
%\bibliography{/home/yzolo/TEX/BIB/fb,/home/yzolo/TEX/BIB/jj,/home/yzolo/TEX/BIB/sc,/home/yzolo/TEX/BIB/books,/home/yzolo/TEX/BIB/el,/home/yzolo/TEX/BIB/bec,/home/yzolo/TEX/BIB/math,/home/yzolo/TEX/BIB/db}

\end{document}